\documentclass[11pt,review,1p,number,sort&compress,a4paper]{elsarticle}
\usepackage{dcolumn}
\usepackage{floatrow}

\usepackage[utf8]{inputenc}
\usepackage{graphics,graphicx}
\usepackage{epstopdf}
\usepackage{float}
\usepackage{natbib}
\usepackage{fancyvrb}
\usepackage{subfigure}
\usepackage[para]{threeparttable}
\usepackage{url}
\usepackage{color}
\usepackage{longtable}

\DeclareFloatFont{tiny}{\scriptsize}
\newcommand{\ea}{{\it et al}}
\newcommand{\cm}{cm$^{-1}$}

\newcommand{\ai}{\textit{ab initio}}

\newcommand{\eqref}[1]{(\ref{#1})}

\newcommand{\p}{^\prime}
\newcommand{\pp}{^{\prime\prime}}

\newcommand{\hto}{\hspace{2mm}}
\newcommand{\hsn}{\hspace{7mm}}
\newcommand{\hsx}{\hspace{6mm}}

\journal{J. Molec. Spectrosc.}

\begin{document}
\begin{frontmatter}

\title{The ExoMol database: molecular line lists for exoplanet and other hot
atmospheres}
\author{Jonathan Tennyson\footnote{Corresponding author}}
\ead{j.tennyson@ucl.ac.uk}
\author{Sergei  N. Yurchenko}
\author{Ahmed F. Al-Refaie}
\author{Emma J. Barton}
\author{Katy L. Chubb}
\author{Phillip A. Coles}
\author{S. Diamantopoulou}
\author{Maire N. Gorman}
\author{Christian Hill}
\author{Aden Z. Lam}
\author{Lorenzo Lodi}
\author{Laura K. McKemmish}
\author{Yueqi Na}
\author{Alec Owens}
\author{Oleg L. Polyansky}
\author{Tom Rivlin}
\author{Clara Sousa-Silva}
\author{Daniel S. Underwood}
\author{Andrey Yachmenev}
\author{Emil Zak}
\address{Department of Physics and Astronomy, University College London, London,
WC1E 6BT, UK}

\date{\today}

\begin{abstract}

  The ExoMol database (www.exomol.com) provides extensive line lists
  of molecular transitions which are valid over extended temperatures
  ranges. The status of the current release of the database is
  reviewed and a new data structure is specified. This structure
  augments the provision of energy levels (and hence transition
  frequencies) and Einstein $A$ coefficients with other key properties,
  including lifetimes of individual states, temperature-dependent
  cooling functions, Land\'e $g$-factors, partition functions, cross
  sections, $k$-coefficients and transition dipoles with phase relations.
Particular
  attention is paid to the treatment of pressure broadening
  parameters. The new data structure includes a definition file which
  provides the necessary information for utilities accessing ExoMol through its
  application programming interface (API). Prospects for the inclusion
  of new species into the database are discussed.
\end{abstract}
\begin{keyword}
infrared \sep visible \sep Einstein $A$ coefficients \sep transition frequencies
\sep partition
functions \sep cooling functions \sep lifetimes \sep cross sections \sep $k$
coefficients
\sep Land\'e $g$-factors
\end{keyword}

\end{frontmatter}

\newpage

\section{Introduction}

Hot molecules exist in many environments in space including cool stars
\cite{09Bexxxx.exo}, failed stars generally known as brown dwarfs
\cite{aha97,09Bexxxx.exo} and exoplanets \cite{13TiEnCo.exo}.  The
atmospheric properties of these objects are known to be strongly
influenced by the spectra of the molecules they contain. On Earth,
spectra of hot molecules are observed in flames
\cite{14BoWeHy,15CoLixx}, discharge plasmas \cite{02ChWaLi.SO,15BiCrGl},
explosions \cite{11CaLiPi.H2O} and in the hot gases emitted, for
example, from smoke stacks \cite{12EvFaCl.CO2}. In addition, high-lying
states can be important in non local thermodynamic equilibrium (non-LTE)
environments both in space, for example, emissions observed from comets
\cite{jt330,jt349}, and on Earth. The spectra of key atmospheric
molecules at room temperature have been the subject of systematically
maintained databases such as HITRAN \cite{jt350,jt453,jt546} and GEISA
\cite{jt504,jt636}. As will be amply illustrated below, the spectra
of hot molecules contain many, many more transitions and so far
attempts to compile systematic databases have been limited.  Databases
of hot molecular spectra do exist for other specialized applications,
such as the EM2C database for combustion applications \cite{12RiSoxx}
or one due to Parigger \ea\ for studies of laser-induced plasmas
\cite{15PaWoSu}.

Planets and cool stars share some common fundamental characteristics:
they are faint, their radiation peaks in the infrared and their
atmosphere is dominated by strong molecular absorbers. Modelling
planetary and stellar atmospheres is therefore difficult as their
spectra are extremely rich in structure with hundreds of thousands to
many billions of spectral lines which may be broadened by
high-pressure and temperature effects.

The ExoMol project \cite{jt528} aims to provide the molecular line
lists that astronomers need in order to understand the physics and chemistry of
astronomical bodies cool enough to form molecules in their
atmospheres. In particular these data are needed for extrasolar planets, brown dwarfs and
cool stars \cite{jt143,aha97,13TiEnCo.exo}, as well as circumstellar
structures such as planetary envelopes and molspheres \cite{08Tsuji}.
In practice these data are also useful for a wide range of other
scientific disciplines; examples include studies of the Earth's
atmosphere \cite{14ScGaHe,15LaPoTs}, hypersonic non-equilibrium
\cite{15LoSuBo}, analysis of laboratory spectra
\cite{14FoGoHe,jt616,16NiReTaKa.CH4}, measurements of hot reactive
gases \cite{13GrFaNi} and the proposed remote analysis of molecular
composition using laser oblation \cite{15TaCa}.

The original ExoMol data structure was very focused in its goal, with the scope
of the data limited to generating lists of transitions \cite{jt548}.
However, it has become obvious that the potential
applications of the ExoMol spectroscopic data are much more diverse. For example,
the ExoMol data can be used  to compute
partition functions \cite{jt571}, cross sections \cite{jt542},
lifetimes \cite{jt624}, Land\'e $g$-factors \cite{jt655} and other properties.
Our aim is to systematically provide this additional data to
maximise its usefulness. To do this requires significant extension of the ExoMol
data
structure, which is the major purpose of this paper. At the same time
this implementation should facilitate the adoption of an application
programming interface (API) between the database and programs using
it. Similar enhancements are actively being pursued by other related
databases such as HITRAN \cite{jt559,HAPI,Honline}. A major new
feature is the inclusion, albeit at a fairly crude level, of
pressure-broadening parameters. These have been shown to be important
for models of exoplanets \cite{jt521,16HeMaxx} and are known to be vital for
many other applications.  Section 2 summarizes current data coverage
of the ExoMol database.  Section 3 reviews the individual line lists
included in the database with particular emphasis on any changes made
to them since they were originally published. Section 4 describes the
expansion of data that is now provided by the ExoMol database. Section
5 gives a formal description of the new data structures implemented to
cover this provision and to provide the API functionality.  Section 6
briefly describes the utility programs available as part of the ExoMol
database and Section 7 the ExoMol website.
Finally we discuss future prospects for the database.

\section{Database coverage}

The ExoMol project aims at complete coverage of the spectroscopic
properties of molecules which are deemed to be important in hot
astrophysical environments. Coverage concerns (a) the molecular
species considered, including isotopologues; (b) the frequency range
considered and (c) the upper temperature range for which the data is
reasonably complete. Both the required temperature and frequency range
completeness are to some extent a judgement on what is required for
astronomical and other studies. For example, a molecule like nitric
acid (HNO$_3$) will dissociate at a relatively low temperature so
coverage above about 500 K is unlikely to be important. On the other
hand, several diatomics such SiO and CO are known to feature in
stellar spectra, and coverage to temperatures over 6000 K is
necessary.

The general ExoMol approach is molecule-by-molecule.
That is, a comprehensive line list is created for a particular molecule
which is then made available in the database.
For more challenging larger systems such as NH$_3$, PH$_3$ and
HNO$_3$, it has been our policy to produce initial, room-temperature
line lists, see \cite{jt466,jt556,jt603} respectively. This allows us
to improve the model, validate the available experimental data and, in
some cases, make spectral assignments.  In all instances, the subsequent
hot line lists \cite{jt500,jt592,jt614} are both more complete and
more accurate and should therefore be used even for studies at low
temperatures.

Thus far ExoMol has not considered ultraviolet (UV) absorption;
however, there is increasing interest in the consequences of UV
radiation on exoplanets \cite{14FoBiLa,14LoPaFr,13BeBaxx} so this may
need to be reviewed in future.  As discussed in the comments below,
molecules are still being added to the database so that coverage by
species is steadily increasing.

Tables \ref{tab:exomoldata} and \ref{tab:otherdata} summarize the
molecules for which the ExoMol database currently provides data. The
division is between species that have explicitly been studied as part
of the ExoMol project (Table \ref{tab:exomoldata}) and those for which
the data has been taken from other studies (Table
\ref{tab:otherdata}). The ExoMol project has a specific methodology
based on the use of spectroscopically-determined potential energy
surfaces and {\it ab initio} dipole surfaces, which are combined with
explicit variational treatments of the nuclear motion problem. For
open shell systems these treatments involve explicit inclusion of
spin-orbit and related curve-coupling effects; the project has
developed a nuclear motion program, {\sc Duo}, for treating coupled
diatomic curves \cite{jt609}.  For closed shell molecules a range of
codes \cite{lr07,jt338,jt339,07YuThJe.method,15YaYuxx.method,jt588}
are used. These are all essentially based on finding near-exact
solutions of the ro-vibrational Schr\"odinger equation for a given
potential energy surface but the level of approximation increases with
the size of the molecule considered.

Data from other sources arise from a variety of methodologies which
range from completely {\it ab initio}, appropriate for systems with
very few electrons, to largely empirical. Table \ref{tab:otherdata}
gives a pointer to the method used in each case; for full details the
reader should consult the cited reference.

\begin{table}[h]
\centering
\caption{Datasets created by the ExoMol project and included in the ExoMol
database.}
\label{tab:exomoldata}
\begin{tabular}{lcrcrll}
\hline\hline
Molecule&$N_{\rm iso}$&$T_{\rm max}$&$N_{elec}$&$N_{\rm lines}$
$^a$&DSName&Reference\\
\hline
BeH&1&2000 &1&16~400&Yadin& Yadin \ea\ \cite{jt529}\\
MgH&3 &2000 &1&10~354&Yadin& Yadin \ea\ \cite{jt529}\\
CaH&1 &2000 &1&15~278&Yadin& Yadin \ea\ \cite{jt529}\\
SiO&5&9000&1& 254~675&EJBT&Barton \ea\ \cite{jt563}\\
HCN/HNC&2$^a$&4000&1&399~000~000&Harris& Barber \ea\ \cite{jt570}\\
CH$_4$&1&1500&1&9~819~605~160&YT10to10& Yurchenko \&\ Tennyson \cite{jt564}\\
NaCl&2&3000&1& 702~271 &Barton&Barton \ea\ \cite{jt583}\\
KCl&4&3000&1& 1~326~765  &Barton&Barton \ea\ \cite{jt583}\\
PN&2&5000&1&142~512&YYLT&Yorke \ea\ \cite{jt590}\\
PH$_3$&1&1500&1&16~803~703~395&SAlTY& Sousa-Silva \ea\ \cite{jt592}\\
H$_2$CO&1&1500&1&10~000~000~000&AYTY& Al-Refaie \ea\ \cite{jt597}\\
AlO&4&8000&3&4~945~580&ATP& Patrascu \ea\ \cite{jt598}\\
NaH&2&7000&2&79~898&Rivlin&Rivlin \ea\ \cite{jt605}\\
HNO$_3$&1&500&1&6~722~136~109&AlJS&Pavlyuchko \ea\ \cite{jt614}\\
CS&8&3000&1&548~312&JnK&Paulose \ea\ \cite{jt615}\\
CaO&1&5000&5&21~279~299&VBATHY&Yurchenko \ea\ \cite{jt618}\\
SO$_2$&1&2000&1&1~300~000~000&ExoAmes& Underwood \ea\ \cite{jt635}\\
\hline
\hline
\end{tabular}

$N_{\rm iso}$ Number of isotopologues considered;\\
$T_{\rm max}$ Maximum temperature for which the line list is complete;\\
$N_{elec}$ Number of electronic states considered;\\
$N_{\rm lines}$  Number of lines: value is for the main isotope.\\
$^a$ A line list for H$^{13}$CN/HN$^{13}$C due to Harris \ea\ \cite{jt447} is
also available.
\end{table}

\begin{table}[h]
\centering
\caption{Datasets not created as part of the ExoMol project but included in the
ExoMol database.}
\label{tab:otherdata}
\small
\begin{tabular}{lcrcrlll}
\hline\hline
Molecule&$N_{\rm iso}$&$T_{\rm max}$&$N_{elec}$&$N_{\rm
lines}$&DSName&Reference&Methodology\\
\hline
H$_3^+$&2$^a$&4000&1&3~070~571&NMT&Neal \ea\ \cite{jt181}&ExoMol\\
H$_2$O &2$^b$&3000&1&505~806~202&BT2& Barber \ea\ \cite{jt378}&ExoMol\\
NH$_3$&2$^c$&1500&1&1~138~323~351&BYTe& Yurchenko \ea\ \cite{jt500}&ExoMol\\
HeH$^+$&4&10000&1&1~431&Engel&Engel \ea\ \cite{jt347}&Ab initio\\
HD$^+$&1&12000&1&10~119&CLT&Coppola \ea\ \cite{jt506}&Ab initio\\
LiH&1&12000&1&18~982&CLT&Coppola \ea\ \cite{jt506}&Ab initio\\
LiH$^+$&1&12000&1&332&CLT&Coppola \ea\ \cite{jt506}&Ab initio\\
ScH&1&5000&6&1~152~827&LYT&Lodi \ea\ \cite{jt599}&Ab initio\\
MgH&1&&3& 30~896&13GhShBe&GharibNezhad \ea\ \cite{13GhShBe.MgH}&Empirical\\
CaH&1& &2&6000 &11LiHaRa& Li {\it et al} \cite{11LiHaRa.cah}&Empirical\\
NH&1&&1&10~414&14BrBeWe&Brooke \ea\ \cite{14BrBeWe.NH}&Empirical\\
CH&2&&4&54~086&14MaPlVa&Masseron \ea\ \cite{14MaPlVa.CH}&Empirical\\
CN&1&&1&195~120&14BrRaWe&Brooke \ea\ \cite{14BrRaWe.CN}&Empirical\\
CP&1&&1&28~735&14RaBrWe&Ram \ea\ \cite{14RaBrWe.CP}&Empirical\\
HCl&1&&1&2588&11LiGoBe&Ram \ea\ \cite{11LiGoBe.HCl}&Empirical\\
CrH&1&&2&13~824&02BuRaBe&Burrows \ea\ \cite{02BuRaBe.CrH}&Empirical\\
FeH&1&&2&93~040&10WEReSe&Wende \ea\ \cite{10WEReSe.FeH}&Empirical\\
TiH&1&&3&181~080&05BuDuBa&Burrows \ea\ \cite{05BuDuBa.TiH}&Empirical\\
\hline
\hline
\end{tabular}

$N_{\rm iso}$ Number of isotopologues considered;\\
$T_{\rm max}$ Maximum temperature for which the line list is complete;\\
$N_{elec}$ Number of electronic states considered;\\
$N_{\rm lines}$  Number of lines: value is for the main isotope.\\
$^a$ There is a H$_2$D$^+$ line list available from Sochi and Tennyson
\cite{jt478}.\\
$^b$ The VTT line list for HDO due to Voronin \ea\ \cite{jt469} is also
available.\\
$^c$ There is a room temperature $^{15}$NH$_3$ line list due to Yurchenko
\cite{15Yurche.NH3}.\\

\end{table}

The molecules listed in Tables \ref{tab:exomoldata} and
\ref{tab:otherdata} do not at present provide a comprehensive set of
species. A number of key species are available from other sources.
HITEMP \cite{jt480} in principle provides a source of data for CO,
OH, NO and CO$_2$. In practice there are new hot line lists for CO
\cite{15LiGoRo.CO}, OH \cite{16BrBeWe.OH}, and CO$_2$
\cite{11TaPe.CO2,14HuGaFr.CO2} which are more recent than the data
given in the current release of HITEMP. High quality line lists for
ozone are also available from elsewhere \cite{13BaMiSt.O3}. Line lists
for some other missing species may also be found in the UGAMOP
(https://www.physast.uga.edu/ugamop/) and Kurucz \cite{11Kurucz.db}
databases. The coverage of these databases for a (possible) exoplanet
characterization mission has recently been reviewed by two of us
\cite{jt578}. VALD3, the latest release of the Vienna atomic line database,
contains line lists for a small number of diatomics \cite{VALD3}.

Finally we note that while the majority of exoplanet spectroscopy is
performed at rather low resolution, very precise spectroscopic data
can be required \cite{13DeBrSn.exo,13BiDeBr.exo,14BrDeBi.exo}, which
is proving an issue for particular key species \cite{15HoDeSn.TiO}.
Therefore, ExoMol aims to be both complete, and as accurate as
possible.

\section{Individual line lists}

 Below we
consider some of the line lists presented in the ExoMol database and listed
in Tables \ref{tab:exomoldata} and \ref{tab:otherdata}. We restrict
our discussion to issues not
covered in the original publication.

One general issue is that Medvedev and co-workers
\cite{highv,15LiGoRo.CO} identified a numerical problem with the
intensities of high overtone transitions computed with the
standard compilation of the diatomic vibration-rotation program {\sc Level}
\cite{lr07}.  Our line lists computed with {\sc Level} have been adjusted
to remove transitions which appeared to have been affected by this issue;
such cases are noted below. The transitions removed are all very weak and
it is anticipated that these changes will have very little effect on practical
applications.
We checked similar calculations performed with our in-house
rovibronic program {\sc Duo} \cite{jt609} and found
similar behaviour: when the value
of the transition dipole moment becomes comparable with
the double precision error ($\approx 10^{-16}$ a.u.), the
corresponding electric dipole intensities essentially represent
numerical noise and have to be removed.

\subsection{MgH}
The ExoMol MgH line list only considers transitions within the
X~$^2\Sigma^+$ ground electronic state \cite{jt529}.
A line list containing A~$^2\Pi$ -- X~$^2\Sigma^+$ and B$^\prime$~$^2\Sigma^+$
-- X~$^2\Sigma^+$
transitions has been given by GharibNezhad \ea\ \cite{13GhShBe.MgH}.
Both line lists are included in the ExoMol database.

Furthermore Szidarovszky and
Cs\'{a}sz\'{a}r \cite{15SzCsxx.MgH} showed that, due to the relatively
low dissociation energy of MgH, consideration of quasi-bound states in
the partition sum significantly alters the partition function at
higher temperatures. However, for reasons of self-consistency we
recommend using the partition function given by Yadin \ea\
\cite{jt529}.

\subsection{CaH}
The ExoMol CaH line list only considers transitions within the
X~$^2\Sigma^+$ ground electronic state \cite{jt529}. A line list containing
E~$^2\Pi$ -- X~$^2\Sigma^+$ transitions has been given by
Li {\it et al} \cite{11LiHaRa.cah}.  Both line lists are included
in the ExoMol database.

\subsection{SiO}

The original SiO ExoMol line lists of Barton \ea\ \cite{jt563} computed using
{\sc Level}
have been truncated by removing transitions with $\Delta v \geq 6$.

\subsection{HCN/HNC}

The combined HCN/HNC Exomol line list of Barber \ea\ \cite{jt570} used
a calculated energy separation (isomerization energy) of 5705 \cm\
between the ground states of HCN and HNC. Recent work by Nguyen \ea\
\cite{15NgBaRu.HCN} suggests that this value is too high.  We have
done some additional \ai\ calculation using {\sc MOLPRO}
\cite{12WeKnKn.methods} at multi-reference configuration interaction
(MRCI) level of theory.  The complete basis set extrapolation (CBS)
value CBS[56]z was found to equal 5356 \cm . Substracting from
this value the zero point energy difference between HCN and HNC wells
of 88 \cm,  gives the value for
the energy separation of 5268 \cm, consistant with the value of 5236 $\pm$
50 \cm\ recommended by Nguyen \ea \cite{15NgBaRu.HCN}.
This value has been adopted by revising the
states file and calculated partition function.

\subsection{CH$_4$}

Although the YT10to10 line list \cite{jt564} represents a major step forward in
the
modelling of hot methane spectra \cite{jt572}, it does not represent a
complete solution to the problem.  Rey \ea\ \cite{14ReNiTy.CH4} have
also produced a line list, using similar procedures to those adopted
by ExoMol, valid to higher temperatures (2000 K) but covering a more
limited spectral range. The temperature coverage can be checked using
available, high-temperature partition functions
\cite{08WeChBo.CH4,15NuKrRe.CH4}.

Yurchenko \ea\ \cite{jt708} have extended the
YT10to10 line list to higher temperatures (2000 K) and improved the
predicted frequencies by replacing computed energy levels with empirical
ones provided by Boudon \cite{14Boudon.CH4}. The resulting line list
has 35 billion transitions and Yurchenko \ea\ propose that most of the lines
in the line list  can be represented by background cross sections.
These temperature-dependent, but
pressure-independent cross sections, generated using the vast majority of the
lines, can then be used to supplement a reduced line list of 203 million
lines selected to be the strongest in each spectral region. If this
methodology proves successful, it will be adopted by the ExoMol
database for other very extensive line lists in future releases. At present
these data are not included in the database.

\subsection{NaCl}

The original NaCl ExoMol line lists of Barton \ea\ \cite{jt583} computed using
{\sc Level} have been truncated by removing transitions with $\Delta v
\geq 8$.

\subsection{KCl}

The original KCl ExoMol line lists of Barton \ea\ \cite{jt583} computed using
{\sc Level} have been truncated by removing transitions with $\Delta v
\geq 12$.

\subsection{PN}

The original PN ExoMol line lists of Yorke \ea\ \cite{jt590} computed using
{\sc Level} have been truncated by removing transitions with $\Delta v
\geq 6$.

\subsection{AlO}

A number of ExoMol users pointed out that the vibrational labels
used in the original AlO line list were not the best ones. Our
analysis of AlO state lifetimes \cite{jt623} reinforced this
impression. The \texttt{.states} file has therefore been updated with
revised vibrational state labels. We note that these labels are only
approximate quantum numbers and this change should not affect any
results obtained with the line list.

\subsection{H$_3^+$}

The H$_3^+$ line list of Neal \ea\ \cite{jt181} (NMT) is the oldest in
the ExoMol database.  While it has continued to demonstrate its
accuracy and predictive power \cite{jt512,jt587}, perhaps surprisingly
so, there are other issues with it.  The line list was constructed
using Jacobi coordinates which do not allow for a full treatment of
the symmetry.  This led to the use of approximate nuclear spin
statistical weights and, more problematically, to the removal of very
weak transitions. Most of these transitions are symmetry-forbidden and
should have zero dipole intensity. However ``forbidden'' rotational
transitions \cite{86PaOk.H3+,jt72} have proved important for both
astrophysical \cite{02GoMcGe.H3+,05OkGeGo.H3+} and laboratory
\cite{jt306,jt340} studies. Methods are available which allow for the
proper treatment of symmetry \cite{jt358,03ScAlHia.H3+}. A new line
list \cite{jt666} is nearly complete which extends the range of the NMT
line list.
The upper energy limit is increased to 25 000 \cm\ and the highest rotational
quantum number state considered in the calculations is  $J=40$.
This line list will remove the symmetry problem as well as improve
both the accuracy and coverage for H$_3^+$, thus making it useful for
both higher and lower temperatures.

\subsection{H$_2$O}
The BT2 line list \cite{jt378} has been outstandingly successful: it
was used for the original detection of water in an exoplanet
atmosphere \cite{jt400}, forms the basis of the well-used BT-Settl
brown-dwarf model \cite{12AlHoFr} and provides the hot water line list
for HITEMP \cite{jt480}. However, although it is more complete, for
many levels it is less accurate than the Ames line list of Partridge
and Schwenke \cite{97PaScxx.H2O}.  Since the construction of BT2 there
has been significant progress in deriving experimental energy levels
for H$_2$$^{16}$O \cite{jt539,jt562}, improvements in the {\it ab
  initio} dipole moment surface \cite{jt509} which have been used to
construct high accuracy room temperature line lists \cite{jt522} and
further improvements in representing the potential energy surface of
the molecule \cite{jt519}. A new H$_2$$^{16}$O line list, named
POKAZATEL, \cite{jt734} which builds on these advances will be
released soon.  The completeness of this line list is illustrated by
the following figures: The highest rotational quantum number used in
the calculations of the wavefunctions was $J=72$ --- close to the highest
$J$ for wich the bound states still do exist. The highest upper energy
levels limit is equal to 40 000 \cm, which is also close to the
dissociation limit of water \cite{jt549}.

While the BT2 and Ames line lists are seriously incomplete for
temperatures above 3000 K, POKAZATEL considers every ro-vibrational
transition in the molecule and therefore is appropriate for studies at
higher temperatures.

Line lists for hot H$_2$$^{17}$O and H$_2$$^{18}$O will also be
released very shortly \cite{jt627}. Again these will take advantage of
the available set of experimental energy levels \cite{jt454,jt482}
obtained using the MARVEL procedure \cite{jt412,12FuCsi.method}.
These line lists will be as complete as BT2 and consider levels with $J
\leq 50$ and energies up to 30 000 \cm.

\subsection{NH$_3$}

Ammonia is a ten electron system like water, so in terms of electronic
calculations one might expect similar improvements here too.
$^{14}$NH$_3$ has also been the subject of a systematic study of
experimental energy levels \cite{jt608} and improvements in its {\it ab
  initio} treatment \cite{11HuScLe.NH3,jt634}. New
experimental data characterizing higher-lying energy levels is also
available
\cite{jt616,jt633}. Again the Ames group have performed high accuracy
studies on this system \cite{11HuScLe.NH3,11HuScLe2.NH3,12SuBrHu.NH3},
although there is no corresponding line list. Work is therefore in
progress to compute a new ammonia line list which will replace
the existing BYTe line list \cite{jt500};
in particular this work the option of using the improved nuclear motion
capability of the program {\sc TROVE} working in curvilinear coordinates
\cite{15YaYuxx.method}. Recent, excellent \ai\ calculations
\cite{jt634} which reproduce very highly excited energy levels of
ammonia up to 18 000 \cm, will be used as the starting point for a
semi-empirical fit of the PES. This should give a line list which is
both more accurate and more extensive than BYTe.

\subsection{CrH}

Chromium hydride is an important astronomical molecule for a variety
of reasons, including classification of brown dwarfs
\cite{99KiReLi.CrH,jt430} and measurements of magnetic fields
\cite{13KuBexx.CrH}.  The ExoMol database currently provides a
substantially empirical line list due to Burrows \ea\
\cite{02BuRaBe.CrH}; we are in the process of computing a more
complete CrH line list covering 8 states and the 4 main isotopologues
\cite{jtCrH}.

\subsection{TiH}

The current titanium hydride line list due to Burrows \ea\
\cite{05BuDuBa.TiH} is in the process of being updated with a more
extensive ExoMol one.

\subsection{Other species}

Besides the species listed above, line lists are in an advanced stage
of construction for VO \cite{jt621}, H$_2$O$_2$ \cite{jt638} (see Al-Refaie  \ea\
\cite{jt620} for a preliminary, room temperature version), SO$_3$ \cite{jt641} (see also Underwood \ea\
\cite{jt554}), C$_2$H$_4$,
CH$_3$Cl (see also Owens \ea\ \cite{jt612}), C$_2$ (see also Furtenbacher \ea\ \cite{jt637}), 
SiH, NS, NO, NaO, AlH and SH
\cite{jt732},
MnH, and PO, PS and PH \cite{jt703}.

Finally we note that work is in progress looking at TiO \cite{jtTiO}.
TiO is a major absorber in cool stars
\cite{00AlHaS1.TiO,13DaKuPl.TiO}. There are TiO line lists available
from Schwenke \cite{98Scxxxx.TiO}, Plez \cite{98Plxxxx.TiO} and VALD3 \cite{VALD3}.
However it would appear that these line lists are incomplete; in
particular there are missing bands that have been observed in the
laboratory \cite{99RaBeDu.TiO}. Furthermore, recent exoplanet studies
suggest that the line frequencies are too inaccurate for high
resolution studies \cite{15HoDeSn.TiO}.

\section{Data Provided}
The original aim of the ExoMol project was the provision of extensive
line lists of energy levels, and hence transition frequencies, and
Einstein $A$ coefficients from which transition intensities and related
information can be computed. The original ExoMol data structure \cite{jt548}
provided precisely for these data in a concise manner.
The demands on and utility of the data provided by the project has led us
to expand the scope of the database. In this section we outline
the new data provision and in the following sections we formally
define the new data structures employed for this purpose.

\subsection{Extended states file}

There are other properties of a molecular state that can be computed
by ExoMol which we now propose to store. The first of these is the
radiative lifetime of each individual state, which can be computed in
a straightforward fashion from the  Einstein $A$ coefficients
\cite{jt624}. Such lifetimes have been shown to be important in both
laboratory \cite{jt306,jt340} and astronomical
\cite{02GoMcGe.H3+,05OkGeGo.H3+} environments. Indeed the radiative
lifetime of a state is a key component in determining the critical
density of species in that state and hence whether it exists in local
thermodynamic equilibrium (LTE) in a given environment. Our recent study of
lifetimes based on use of ExoMol data suggested that methane should
show particularly interesting lifetime effects \cite{jt624}.
These computed lifetimes can be compared with experimental measurements,
where available, providing a check on our calculated dipole moments.
This is particularly useful for rovibronic spectra, where
measurements of absolute
transition intensities are unusual.

The second state-dependent property are Land\'e $g$-factors. These provide
the behaviour of the states in the presence of a weak magnetic field
as given by the Zeeman effect.
Molecular spectra can provide important information on magnetic
fields in stars, brown dwarfs and  exoplanets \cite{13KuBexx.CrH,15AfBexx}.
For open shell diatomics, Land\'e $g$-factors are given by a fairly
straightforward formula in terms of standard Hund's case (a) quantum
numbers \cite{02BeSoxx.diatom}.
The vibronic wavefunctions computed by {\sc Duo} \cite{jt609}
contain the necessary information to compute these  $g$-factors and
this has currently recently been implemented within {\sc Duo}  \cite{jt655}.
The magnetic fields in some astronomical environments are strong
enough that the Paschen-Back effect becomes important
\cite{13KuBexx.CrH,07ShFlBe.CN}. In this case
the shifts in the energy levels do not have a simple dependence on the
magnetic field \cite{05BeBrFl.diatom} and they will have to be explicitly
computed on a case-by-case
basis. Land\'e $g$-factors are only given for systems whose
spectroscopic model includes open shell electronic states (ie ones
with unpaired electrons) since these are the systems
whose states which show
significant splitting in a magnetic field.

Besides (approximate) quantum numbers, which may appear in several
forms, it is also often desirable to store more than one estimate of
the energy level of the state. A general methodology, known as MARVEL,
 is available for
obtaining self-consistent sets of empirical energy levels from
networks of transitions \cite{jt412,12FuCsi.method}, which means that
sets of empirical energy levels are available for key molecules
\cite{jt454,jt482,jt539,jt576,13FuSzMa.H3+,jt637}. In this case it is
preferable to replace computed energy levels with empirical ones, as has
already been done for the HCN/HNC \cite{jt570} and the CS \cite{jt615} line
lists. It is also recommended to keep the original computed levels as
well as the uncertainties which are generally available for the
empirical levels. The inclusion of approximate quantum numbers and extra energy levels included
has been decided on a molecule-by-molecule basis; information on this is
contained
in the isotopologue definition file (see Section 5.3).
\clearpage

\subsection{Cross sections,  $k$-coefficients and pressure effects}

Many of the line lists stored in the database are huge. This makes
finding other forms of storing the data desirable.  ExoMol data has
been used to generate temperature-dependent cross sections as a
function of wavenumber \cite{jt542}. Such cross sections now form part
of the database which will also be expanded to contain
pressure-dependent cross sections and $k$-coefficient tables generated from
them.

Cross sections depend on pressure, as well as temperature, due to
collisional broadening by the species in the atmosphere. The provision
of temperature-, pressure- and broadening species-dependent cross
sections would be both computationally very demanding and again result
in very large datasets. Instead we have chosen to primarily provide parameters
which characterise pressure profiles for key species plus software
allowing cross sections to be generated for selected
temperature-pressure parameters on the user's own computer.  There is
a widespread recognition that for detailed studies, such as those
performed in the Earth's atmosphere at high resolution, Voigt profiles
only provide an approximate solution to the problem \cite{jt584}.
However Voigt profiles are in widespread use and are easily computed
\cite{79Humlic}; they are therefore used to represent pressure broadening
effects in ExoMol.

The new release of ExoMol includes, where possible, pressure
broadening parameters. So far, these have not been considered for
vibronic spectra. HITRAN provides a source of air-broadening and
self-broadening parameters \cite{07GoRoGa.broad} so in the absence of
other sources these are taken from HITRAN 2012 \cite{jt557}.  In gas
giants H$_2$ and He are the major broadeners. There is no systematic
source of parameters for these broadeners so we have to treat each
system on a case-by-case basis. While some work has been performed for
broadening of hot water by H$_2$ and He \cite{jt544,jt636}, more data
would clearly be helpful in this area. Hedges and Madhusudhan
\cite{16HeMaxx} recently put this in context by quantifying the
effects of various factors involved in modelling line broadening in
exoplanetary atmospheres, including completeness of broadening
parameters, on molecular absorption cross sections.

ExoMol will provide Lorentzian half-widths and their
temperature-exponents for key molecule-broadener systems. The
availability of these parameters varies with species and broadener. In
any case there are not individual values for every molecular line and
often parameters have only been measured or calculated for a small
fraction of transitions. Therefore, rather than pre-determine values
for every molecular line and increase the size of the line lists,
which are very large for some molecules, a separate pressure
broadening parameters file is provided.  This file contains three
types of parameters: experimental, theoretical and semi-empirical.
Experimental, theoretical and semi-empirical parameters from the
literature are presented with their respective full or partial quantum
number assignments.  Additional semi-empirical parameters are
determined by compiling all experimental and theoretical parameters as
a function of $J\pp$, the total rotational quantum number of the lower
level of the transition, and computing an average value for each
$J\pp$. To avoid introducing additional error, no extrapolation is
attempted beyond the $J\pp_{\rm max}$ for which data are available,
the parameters are simply assumed to be constant from this point.
This is a very basic model similar to that used by other line-width
studies \cite{97BuMaHu.broad,jt483,14AmBaTr.broad}. Constructing the
pressure broadening parameters file in this way ensures that
parameters are provided for every spectral line. Table
\ref{tab:broad_ref} lists the main sources used to provide pressure
broadening parameters.  Clearly further and improved parameters would
be welcome.

\begin{longtable}{llll}
\caption{Sources of pressure broadening parameters for key molecule-broadener
species currently being considered by ExoMol. Entries marked $^*$ were
extracted from HITRAN \cite{jt557,16WiGoKo.pb}.
\label{tab:broad_ref}
}\\
\tiny \\
\hline \\
\multicolumn{1}{c}{Molecule}  & Broadener & Reference & Methodology \\
\endfirsthead
\multicolumn{3}{c}%
{\tablename\ \thetable\ -- \textit{Continued from previous page}} \\
\hline
Molecule  & Broadener & Reference & Methodology \\
\hline
\endhead
\hline \multicolumn{3}{r}{\textit{Continued on next page}} \\
\endfoot
\hline
\endlastfoot
\hline
H$_{2}$O & H$_{2}$  & Lavrentieva \ea\ \cite{14LaVoNa.h2opb}        &
Semi-empirical \\
         &          & Lavrentieva \ea\ \cite{14LaDuMa.h2opb}        &
Semi-empirical \\
         &          & Steyert \ea\ \cite{Steyert2004183}            & Experimental
\\
         &          & Brown \&\ Plymate \cite{Brown1996263}         & Experimental
\\
         &          & Brown \ea\ \cite{05BrBeDe.h2opb}              & Experimental
\\
         &          & Gamache \ea\ \cite{Gamache1996471}            & Theoretical
\\
         &          & Dick \ea\ \cite{Dick2009619}                  & Experimental
\\
         &          & Faure \ea\ \cite{jt544}                       & Theoretical
\\
         &          & Langlois \ea\ \cite{Langlois1994272}          & Experimental
\\
         &          & Dutta \ea\ \cite{93DuJoGo.h2opb}              & Experimental
\\
         &          & Golubiatnikov \cite{05Goxxxx.h2opb}           &
Semi-empirical \\
         &          & Zeninari \ea\ \cite{04ZePaCo.h2opb}           &
Semi-empirical \\
         &          & Drouin and Wiesenfeld \cite{12DrWixx.h2opb}   & Theoretical
\\
H$_{2}$O & He       & Lavrentieva \ea\ \cite{14LaVoNa.h2opb}        &
Semi-empirical \\
         &          & Lavrentieva \ea\ \cite{14LaDuMa.h2opb}        &
Semi-empirical \\
         &          & Steyert \ea\ \cite{Steyert2004183}            & Experimental
\\
         &          & Gamache \ea\ \cite{Gamache1996471}            & Theoretical
\\
         &          & Dick \ea\ \cite{Dick2009619}                  & Experimental
\\
         &          & Lazarev \ea\ \cite{95LaPnSu.h2opb}            & Experimental
\\
         &          & Dutta \ea\ \cite{93DuJoGo.h2opb}              & Experimental
\\
         &          & Golubiatnikov \cite{05Goxxxx.h2opb}           &
Semi-empirical \\
         &          & Zeninari \ea\ \cite{04ZePaCo.h2opb}           &
Semi-empirical \\
H$_{2}$O &  Air     & M\'erienne \ea\ \cite{03MeJeHe.h2opb}$^*$         &

Experimental \\
         &          & Gamache \&\ Hartmann \cite{04GaHaxx.h2opb}$^*$    &

Semi-empirical \\
         &          & Gasster \ea\ \cite{88GaToGo.h2opb}$^*$           &
Experimental \\
         &          & Payne \ea\ \cite{08PaDeCa.h2opb}$^*$              &

Experimental \\
         &          & Gamache \cite{05Gaxxxx.h2opb}$^*$                 &

Semi-empirical \\
         &          & Gamache \&\ Laraia \cite{09GaLaxx.h2opb}      &
Semi-empirical \\
H$_{2}$O & H$_{2}$O & M\'erienne \ea\  \cite{03MeJeHe.h2opb}$^*$        &

Experimental  \\
         &          & Gamache \&\ Hartmann \cite{04GaHaxx.h2opb}$^*$    &

Semi-empirical \\
         &          & Markov  \cite{94Maxxxx.h2opb}$^*$              & Experimental
\\
         &          & Golubiatnikov \ea\ \cite{08GoKoKr.h2opb}$^*$      &

Experimental \\
         &          & Cazzoli \ea\ \cite{08CaPuBu.h2opb}$^*$            &

Semi-empirical \\
CH$_{4}$ & H$_{2}$  & Pine \cite{92Pinexx.ch4pb}                    & Experimental
\\
         &          & Margolis  \cite{93Maxxxx.ch4pb}               & Experimental
\\
         &          & Fox \ea\ \cite{88FoJeSt.ch4pb}                & Experimental
\\
         &          & Strong \ea\ \cite{93StTaCa.ch4pb}             & Experimental
\\
CH$_{4}$ & He       & Pine  \cite{92Pinexx.ch4pb}                   & Experimental
\\
         &          & Varanasi \&\ Chudamani\cite{90VaChxx.ch4pb}   & Experimental
\\
         &          & Gabard \ea\ \cite{04GaGrGr.ch4pb}             & Experimental
\\
         &          & Grigoriev \ea\ \cite{01GrFiTo.ch4pb}          &
Semi-empirical \\
         &          & Fox \ea\ \cite{88FoJeSt.ch4pb}                & Experimental
\\
CH$_{4}$ & Air      & Predoi-Cross \ea\ \cite{06PrBrDe.ch4pb}$^*$       &

Experimental        \\
         &          & Smith \ea\ \cite{09aSmBePr.ch4pb}$^*$             &

Experimental         \\
         &          & Brown \ea\ \cite{03BrBeCh.ch4pb}$^*$              &

Semi-empirical     \\
         &          & Anthony \ea\ \cite{08AnNiWr.ch4pb}            & Theoretical
         \\
CH$_{4}$ & CH$_{4}$ & Predoi-Cross \ea\ \cite{05PrBrDe.ch4pb}$^*$       &

Experimental        \\
         &          & Smith \ea\ \cite{09bSmBePr.ch4pb}$^*$             &

Experimental         \\
         &          & Brown \ea\ \cite{03BrBeCh.ch4pb}$^*$              &

Semi-empirical     \\
NH$_{3}$ & H$_{2}$  & Pine \ea\ \cite{93PiMaBu.nh3pb}$^*$               &

Experimental  \\
         &          & Hadded \ea\ \cite{01HaArOr.nh3pb}$^*$             &

Experimental \\
         &          & Sharp \&\ Burrows \cite{07ShBuxx.nh3pb}$^*$      &
Semi-empirical \\
NH$_{3}$ & He       & Pine \ea\ \cite{93PiMaBu.nh3pb}$^*$               &

Experimental \\
         &          & Hadded \ea\ \cite{01HaArOr.nh3pb}$^*$             &

Experimental \\
         &          & Sharp \&\ Burrows \cite{07ShBuxx.nh3pb}$^*$       &

Semi-empirical \\
NH$_{3}$ & Air  & Brown \&\ Peterson \cite{94BrPexx.nh3pb}$^*$      & Experimental
\\
NH$_{3}$ & NH$_{3}$ & Brown \&\ Peterson \cite{94BrPexx.nh3pb}$^*$      &

Experimental \\
PH$_{3}$ & H$_{2}$  & Bouanich \ea\ \cite{Bouanich2004195}          &
Semi-empirical  \\
         &          & Levy \ea\ \cite{Levy1993172}                  & Experimental
\\
         &          & Sergent-Rozey \ea\ \cite{SergentRozey198866}  & Experimental
\\
         &          & Salem \ea\ \cite{Salem200423}                 & Experimental
\\
         &          & Pickett \ea\ \cite{Pickett1981197}            & Experimental
\\
         &          & Levy \ea\ \cite{Levy199420}                   & Experimental
\\
PH$_{3}$ & He       & Salem \ea\ \cite{Salem2005247}                & Experimental
\\
         &          & Levy \ea\ \cite{Levy1993172}                  & Experimental
\\
         &          & Sergent-Rozey \ea\ \cite{SergentRozey198866}  & Experimental
\\
         &          & Pickett \ea\ \cite{Pickett1981197}            & Experimental
\\
         &          & Levy \ea\  \cite{Levy199420}                  & Experimental
\\
PH$_{3}$ & Air  & Butler \ea\ \cite{06BuSaKl.ph3pb}$^*$             & Experimental
\\
         &          & Kleiner \ea\ \cite{Kleiner2003293}$^*$            &

Semi-empirical \\
PH$_{3}$ & PH$_{3}$ & Butler \ea\ \cite{06BuSaKl.ph3pb}$^*$             &

Experimental \\
         &          & Kleiner \ea\ \cite{Kleiner2003293}$^*$            &

Semi-empirical \\
H$_{2}$CO & H$_{2}$ & Nerf \cite{75Nerfxx.hrcopb}                   & Experimental
\\
H$_{2}$CO & He      & Nerf \cite{75Nerfxx.hrcopb}                   & Experimental
\\
H$_{2}$CO & N$_{2}$ & Jacquemart \ea\ \cite{10JaLaKw.hrcopb}        &
Semi-empirical  \\
H$_{2}$CO & H$_{2}$O& Jacquemart \ea\ \cite{10JaLaKw.hrcopb}$^*$        &

Semi-empirical  \\
HCN       & Air & Yang \ea\ \cite{08YaBuGo.hcnpb}$^*$               & Experimental
\\
          &         & Devi \ea\ \cite{04DeBeSm.hcnpb}$^*$               &

Semi-empirical \\
          &         & Risland \ea\ \cite{03RiDeSm.hcnpb}$^*$            &

Semi-empirical \\
HCN       & HCN     & Devi \ea\ \cite{04DeBeSm.hcnpb}$^*$               &

Semi-empirical \\
          &         & Devi \ea\ \cite{03DeBeSm.hcnpb}$^*$               &

Semi-empirical \\
CS        & Air & Blanquet \ea\ \cite{99BaWaBo.cspb}$^*$            &
Semi-empirical  \\
CS        & CS      & Misago \ea\ \cite{09MiLeBo.cspb}$^*$              &

Semi-empirical \\
\hline

\end{longtable}

Another compact
form of data input to exoplanet modelling codes involves the use of the
$k$-coefficient approximation
\cite{92FuLixx.method,95Kratzx.method,96IrCaTa.method},
for example, by the
NEMESIS code  \cite{08IrTeKo.model} and cross sections, as used
by Tau-REx \cite{13HoTeTi.exo,jt593,jt611}.
Temperature- and pressure-dependent $k$-coefficients are also been provided
for
certain key species.

\subsection{Partition and cooling functions}

 In practice
the ExoMol project always required partition functions, the study of which has
often been performed independently \cite{jt169,jt263,jt304,jt571},
and on occasion adopted from other studies \cite{08WeChBo.CH4}.
Temperature-dependent partition functions are now formally included
as part of the data structure.

Of course in various environments such as the early Universe and
regions of star or planetary formation, the transformation
of energy into radiation through molecular emissions provide
an important source of cooling. Temperature-dependent cooling
functions are important \cite{jt489,jt506,jt551}
and can also be computed from ExoMol line lists \cite{jt624}. These
are also now included in the data structure.

\subsection{Dipoles}

While Einstein $A$ coefficients are sufficient for the vast majority
of radiative transport applications, the construction of these loses
the phase information contained in the individual, complex-valued transition
dipole
moments. Incorporating the phase information in the ExoMol database
will make it useful for theoretical modelling of the effects of
an electric field on molecular structure. The areas of application
includes cooling and trapping of molecular beams (see
Ref.~\cite{Lemeshko13} and references therein), manipulating the
long-range molecular interactions and collisional
dynamics~\cite{Campbell09,Lemeshko09,Lemeshko12}, molecular
orientation and alignment~\cite{Boca00,Friedrich00,Lemeshko13}, as
well as rotational
spectroscopy~\cite{Aldegunde08,Aldegunde09,Stuhl12,Garttner13}. As
shown in Section~\ref{dipole_file}, the absolute values of transition dipole
moments can be computed from the Einstein $A$ coefficients and it is
sufficient to complement this with only the sign of transition dipole moment for
each molecular line.

\subsection{Spectroscopic Models}

In addition to the spectroscopic data provided by the ExoMol database,
each ExoMol line list has been constructed from a detailed
spectroscopic model of the given molecule. In general this model
includes a (spectroscopically-determined) potential energy surface and
{\it ab initio} dipole moment surfaces, and input to the appropriate
nuclear motion program. For systems with multiple electronic states,
the spectroscopic model also includes coupling between electronic
states, e.g. spin-orbit, electronic angular momentum etc.

These spectroscopic models are a valuable product of the ExoMol
procedure and there are several reasons for preserving them. Firstly,
the spectroscopic model is ultimately the source of error in the final
line list and is thus important when investigating discrepancies
between experiment or astronomical observations and theoretical
predictions. Secondly, they can be used as a starting point for future
models or line lists. For example, new experimental data can be
incorporated through refinement of potential energy curves; improved
{\it ab initio} methodologies can be incorporated by improvements to
the dipole moment surface, or further electronic states can be added
to the model. Finally, they can be useful for other applications
besides calculation of the original line list. Current work within the
ExoMol group is investigating the sensitivity of spectral lines to a
possible variation in the proton-to-electron mass ratio
\cite{15OwYuTh.NH3,15OwYuPo.H3O+}.

An important part of each spectroscopic model is the input file to the
appropriate nuclear motion program (e.g. {\sc
  Level} \cite{lr07}, {\sc Duo} \cite{jt609},  {\sc DVR3D} \cite{jt338}, {\sc TROVE}
\cite{07YuThJe.method}). For the diatomic codes {\sc
  Level}  and {\sc Duo}, the program input generally includes full
details of the appropriate (potential, dipole and coupling) curves \cite{jt632}.
For the polyatomic nuclear motion programs {\sc DVR3D} and {\sc TROVE},
the PES and DMS are provided as standalone functions, usually written in
Fortran, with the program inputs as separate files.
These spectroscopic models are not part of the formal ExoMol data structure
described in the next section but can be accessed from the webpage of
the appropriate isotopologue in files labelled \texttt{.model}.

\section{Data Structures}

The data structure outlined below represents a significant extension
of the original ExoMol data structure described by Tennyson \ea\
\cite{jt548}.  However the core structure of the two files, the states and
transitions files, used in the original specification remains
unchanged meaning that utilities designed to work with the previous
structure will still work.

\begin{figure}[t]
\centering
\includegraphics[width=0.795\textwidth]{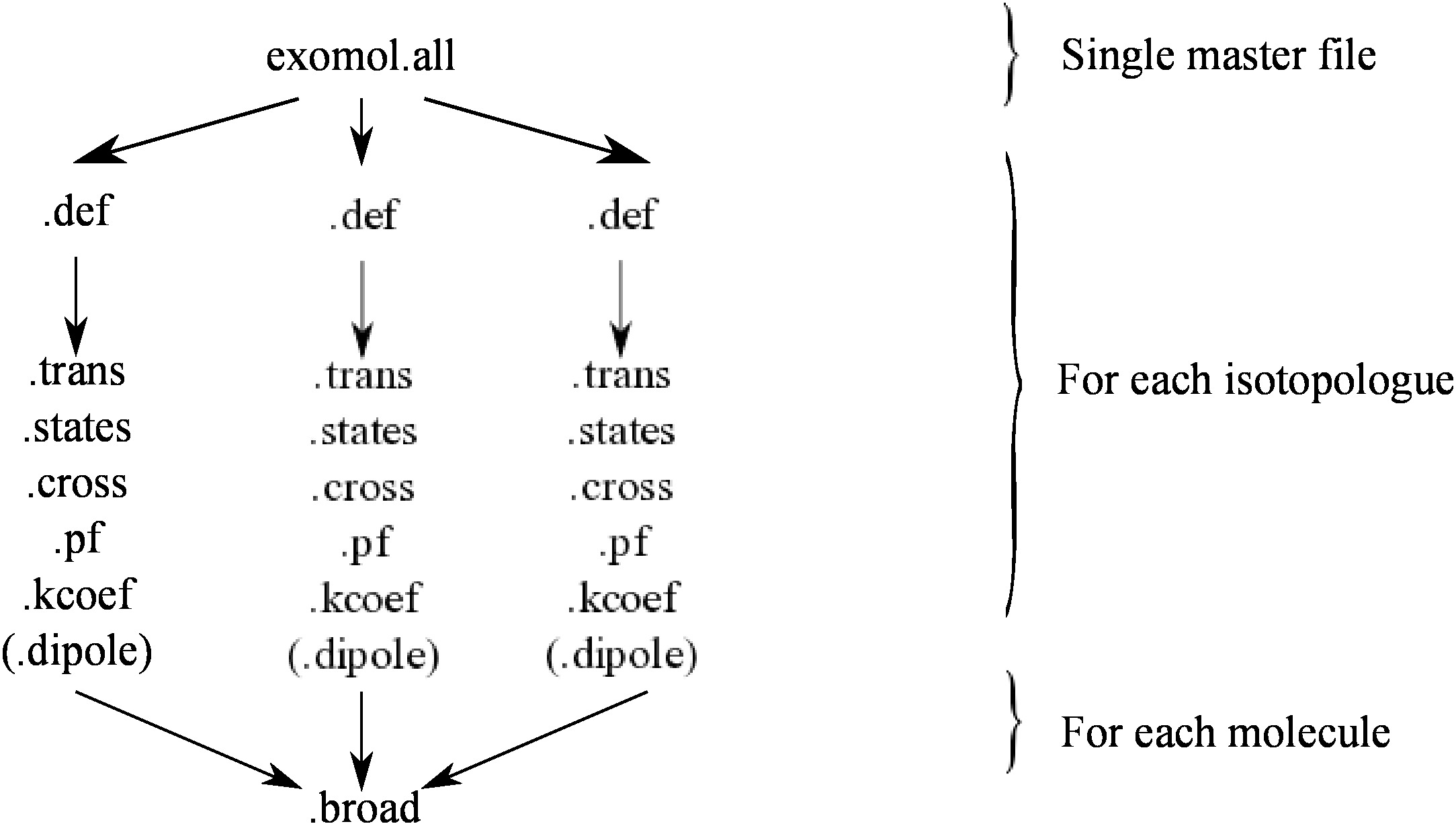}
\caption{Structure of the ExoMol data files. There is a single
master file describing the database, $N_{\rm mol}$ molecules and
$N_{\rm iso}$ files per molecule.}
\label{f:tree}
\end{figure}

A summary of the contents of the database is stored in a master file,
exomol.all.  This file points towards the files which store the actual
data.  Table~\ref{tab:files} gives an overview of these files, which
are available for each isotopologue under the updated ExoMol
structure. Each of these files is specified in turn below.
Figure~\ref{f:tree} gives a schematic representation of the file
structure of the ExoMol data.

\begin{table}
\caption{Specification of the ExoMol file types. (Contents in brackets are
optional.)}
\label{tab:files}
\tiny
\begin{tabular}{lcll}
\hline\hline
File extension & $N_{\rm files}$&File DSname &  Contents\\
\hline
\texttt{.all} &1& Master& Single file defining contents of the ExoMol
database..\\
\texttt{.def} &$N_{\rm tot}$& Definition& Defines contents of other files for
each isotopologue.\\
\texttt{.states} &$N_{\rm tot}$& States & Energy levels, quantum numbers,
(lifetimes), (Land\'e $g$-factors).\\
\texttt{.trans} &$^a$& Transitions & Einstein $A$ coefficients, (wavenumber).\\
\texttt{.broad} &$N_{\rm mol}$& Broadening & Parameters for pressure-dependent
line profiles.\\
\texttt{.cross}&$^b$& cross sections& Temperature or temperature and pressure-dependent
cross sections.\\
\texttt{.kcoef}&$^c$& $k$-coefficients& Temperature and pressure-dependent
$k$-coefficients.\\
\texttt{.pf}& $N_{\rm tot}$&Partition function&   Temperature-dependent
partition function,
(cooling function).\\
\texttt{.dipoles}&$N_{\rm tot}$& Dipoles & Transition dipoles including
phases.\\
\texttt{.overview}&$N_{\rm mol}$& Overview & Overview of datasets available.\\
\texttt{.readme}&$N_{\rm iso}$& Readme & Specifies data formats.\\
\texttt{.model}&$N_{\rm iso}$& Model & Model specification.\\
\hline\hline
\end{tabular}

\noindent
$N_{\rm files}$ total number of possible files.\\
$N_{\rm mol}$ Number of molecules in the database\\
$N_{\rm tot}$ is the sum of $N_{\rm iso}$ for the $N_{\rm mol}$ molecules in the
database\\
$N_{\rm iso}$ Number of isotopologues considered for the given molecule\\
$^a$ There are $N_{\rm tot}$ sets of \texttt{.trans} files but for molecules
with large
numbers of transitions the \texttt{.trans} files are subdivided into wavenumber
regions.\\
$^b$ There are $N_{\rm cross}$ sets of \texttt{.cross} files for
isotopoplogue.\\
$^c$ There are $N_{\rm kcoef}$ sets of \texttt{.kcoef} files for  each
isotopoplogue.\\
\end{table}

\subsection{File naming convention}

The naming convention for the files in the ExoMol database has the
following specification.  All file names, with the exception of the
Pressure Broadening file, have a common part built using the template
\verb!Iso-slug__LineList__!. The \lq\lq Iso-slug" is a machine-readable,
plain ASCII-text \lq\lq XML-safe" unique identifier for the molecular
isotopologue, for example, \verb!14N-1H3! is the iso-slug for
$^{14}$NH$_3$. Ions are denoted by an appended '\verb!_mZ!' or
'\verb!_pZ!' for charges $-Z$ and $+Z$ respectively  
(with the value of Z omitted if it is unity); so \verb!1H3_p!
is the iso-slug for $^1$H$_3^{+}$. \lq\lq Linelist" is the line list publication
name, see the column ``DSName'' in Tables~\ref{tab:exomoldata} and
\ref{tab:otherdata}; for example, the published names for the
$^{14}$NH$_3$ and H$_3^{+}$ line lists are BYTe and NMT respectively.
For external datasets that have been cast in ExoMol format, this
name is an 8 character year-authors string which follows the convention
of the IUPAC Task Group on water spectroscopy \cite{jt454}.
The extensive bibliography files, in BibTeX format, provided  on the ExoMol
website for each molecule in the database, uses the same naming convention.

This common part of the file name can be followed by additional
identifiers specific to different file types. For example, the
transition file \texttt{.trans} for polyatomic molecules is usually
split into smaller wavenumber ranges in order to simplify the
manipulation of data. In this case the \verb!Iso-slug__LineList__! is
followed by \verb!NUMIN-NUMAX!, which specifies
the wavenumber range in \cm. The absorption cross section \texttt{.cross}
file names also contain a wavenumber range, representing the full
coverage of the line list, followed by the temperature in Kelvin,
pressure in bar and the binning interval, or resolution (in \cm), for
which the cross sections have been computed. The file names have the
form \verb!NUMIN-NUMAX__Temperature[K]__Pressure[bar]__resolution!
after the common part.

In the case of the Pressure Broadening file \texttt{.broad}, the file name does not
contain
the string \verb!LineList__!
as the data in the broadening files is not line list specific and has been
compiled from a number of sources.
Here the string \verb!Iso-slug__! is simply followed by \verb!Broadener!
identifying the broadener.
\begin{flushleft}
 For example, the current BT2 line list package consists of the definition file
\linebreak \verb!1H2-16O__BT2.def!,
\linebreak the States file
\linebreak \verb!1H2-16O__BT2.states!,
\linebreak sixty  Transitions files:
\linebreak \verb!1H2-16O__BT2__00000-00500.trans!,
\linebreak \verb!1H2-16O__BT2__00500-01000.trans!,
\linebreak \ldots
\linebreak \verb!1H2-16O__BT2__29500-30000.trans!,
\linebreak Partition function file \verb!1H2-16O__BT2.pf!, 
\linebreak Pressure Broadening files:
\linebreak \verb!1H2-16O__H2.broad!,
\linebreak \verb!1H2-16O__He.broad!,
\linebreak \verb!1H2-16O__air.broad!,
\linebreak \verb!1H2-16O__self.broad!
\linebreak and eighteen cross section files:
\linebreak \verb!1H2-16O__BT2__00000-30000__296K__0bar__0.01.cross!,
\linebreak \verb!1H2-16O__BT2__00000-30000__400K__0bar__0.01.cross!,
\linebreak \ldots
\linebreak \verb!1H2-16O__BT2__00000-30000__900K__0bar__0.01.cross!,
\linebreak \verb!1H2-16O__BT2__00000-30000__1000K__0bar__0.01.cross!,
\linebreak \verb!1H2-16O__BT2__00000-30000__1200K__0bar__0.01.cross!,
\linebreak \ldots
\linebreak \verb!1H2-16O__BT2__00000-30000__3000K__0bar__0.01.cross!,
\end{flushleft}
Here \verb!1H2-16O! is the iso-slug for H$_2{}^{16}$O, \verb!BT2! is the name of
the water line list \cite{jt378},
\verb!00000-00500! is the wavenumber range 0--500~cm$^{-1}$, H$_2$, He, air  and
self are the broadeners, 296 K is the temperature, 0 bar is the pressure
and 0.01 is the binning interval for the cross sections in wavenumbers.

For most cases a single \texttt{.broad} file is provided for a given
molecule and designated for the parent (most abundant) isotopologue.
Should broadening parameters be required for other isotopologues, this
file should be employed.  In a few cases where isotopic substitution
lowers the symmetry, e.g. HDO as compared to H$_2$O, then extra
isotopologue-specific broadening files may be provided, although thus
far there are no such files.

\SaveVerb{term}|exomol.all|
\begin{table}
\caption{Format of the ExoMol master file, \protect\UseVerb{term};
each entry starts on a new line.}
\begin{tabular}{lrll}
\hline\hline
Field &  Fortran Format & C Format &   Description \\
\hline
\multicolumn{4}{l}{{\bf Header Information}} \\
\texttt{ID} & A13  & \%13s & Always the ASCII string ``EXOMOL.master'' \\

$V_{\rm all}$ & I8  & \%8d & Version number, format YYYYMMDD, recording \\

              &      &     & last update of the whole database\\
$N_{\rm mol}$ &  I4 & \%4d & Number of molecules in the database \\
\multicolumn{4}{l}{Followed by $N_{\rm mol}$ entries of the form}
\\
\multicolumn{4}{l}{{\bf Molecule Information}} \\
$N_{\rm name}$ & I3 & \%3d& Number of molecule names listed.\\
\multicolumn{4}{l}{Followed by $N_{\rm name}$ entries of the form}\\
\texttt{MolName} & A27 & \%27s & (Common) name of the molecule\\
\texttt{MolFormula} & A27 & \%27s & Molecule chemical formula\\
$N_{\rm iso}$ & I4 & \%4d& Number of isotopologues considered\\
\texttt{MolKey} &  A27 & \%27s & Inchi key of the isotopologue \\
\texttt{IsoFormula} & A27 & \%27s & Isotopologue chemical formula\\
\texttt{Iso-slug} &  A160 & \%160s & Isotopologue slug identifier, see text for
details\\
\texttt{DSName} &  A10 & \%10s & Isotopologue dataset name\\
$V$ & I8  & \%8d & Version number with format YYYYMMDD\\
Metadata& & & Free format\\
\hline
\end{tabular}
  \label{tab:all}
\end{table}

\subsection{The master file}

The file exomol.all is designed to be machine searchable making it
easy to check the current contents of the database and when any of it
was last updated. The file is structured as a list of molecules
with the isotopologues associated with that molecule. The actual
ExoMol data files are stored by isotopologue. Table~\ref{tab:all}
specifies the format of the file while Table~\ref{tab:alleg} gives a
portion of the current contents for water with two isotopologues,
H$_2$$^{16}$O and HD$^{16}$O. The final entry gives web-searchable
meta-data.

\SaveVerb{term}|exomol.all|
\begin{longtable}{ll}
\hline
\verb!  EXOMOL.master        !&\verb!   # ID                                  !\\
\verb!  20160315             !&\verb!   # Version number with format YYYYMMDD !\\
\verb!    30                 !&\verb!   # Number of molecules in the database!\\
\verb!    1                  !&\verb!   # Number of molecule names listed  !\\
\verb!  water                !&\verb!   # Name of the molecule             !\\
\verb!  H2O                  !&\verb!   # Molecule chemical formula        !\\
\verb!    2                  !&\verb!   # Number of isotopologues considered  !\\
\verb!  XLYOFNOQVPJJNP-DYCDLGHISA-N    !&\verb!   # Inchi key of molecule  !\\
\verb!  (1H)(2H)(16O)            !&\verb!   # Iso-slug                    !\\
\verb!  1H-2H-16O        !&\verb!   # IsoFormula                        !\\
\verb!  VTT                  !&\verb!   # Isotopologue dataset name        !\\
\verb!  20160314             !&\verb!   # Version number with format YYYYMMDD        !\\
\verb!  XLYOFNOQVPJJNP-UHFFFAOYSA-N     !&\verb!   # Inchi key of molecule      !\\
\verb!  (1H)2(16O)              !&\verb!   # Iso-slug                               !\\
\verb!  1H2-16O           !&\verb!   # IsoFormula                            !\\
\verb!  BT2                  !&\verb!   # Isotopologue dataset name             !\\
\verb!  20160220             !&\verb!   # Version number with format YYYYMMDD        !\\
\hline                                                                    
          
\caption{ Extract from the \protect\UseVerb{term} master file showing the three line header and the portion for the
water molecule.}\label{tab:alleg}
 \end{longtable}             
                             
\subsection{The definition file}
                             
A new addition to the ExoMol format is the inclusion of the definition
file with extension \texttt{.def}. The definition file gives information on
what ExoMol provides for a particular isotopologue and describes how
the data can be used. The definition file serves multiple purposes:
\begin{itemize}              
\item \textbf{Standardized ExoMol file usage}:
  The ExoMol format \cite{jt548} \texttt{.states} file contains 4
  standard fields: The ID of the state, the energy, the total
  degeneracy and the $J$ quantum number. However additional fields such
  as the symmetry and vibrational quantum numbers are not standardized
  and differ between molecules. For usage
  that requires these additional quantum numbers, a subroutine or
  function must be created for each molecule to be read. The
  definition file therefore provides a means to overcome this by
  providing a single subroutine with the information to read any states
  file provided by the ExoMol project.
This allows codes to easily integrate the ExoMol format only once and provide
support for all molecules in the ExoMol project without additional changes.

\item \textbf{Improved database function}: The definition file also provides
detailed information on all fields in the ExoMol format. This means that
database functionality such as sorting, filtering, splitting, selecting etc on
specific or multiple fields can be easily performed on single or multiple
molecules.

\item \textbf{Facilitated updates}: Finally, updates of particular molecules can
be easily propagated to any user of the ExoMol format. Each definition file is
assigned a version number; this allows the user to check
whether there are any updates to a particular molecule, with the possibility of
an automatic download of any modified files
and immediate usage without change to codes.
\end{itemize}
The format is defined in Table~\ref{tab:def_format}. The data format
can be subdivided into different sections (highlighted in bold)
describing a different aspect of the molecule in the ExoMol format.
Each section may contain multiple subsections that define further
aspects of that section. The file format therefore
takes a tree structure.  An example definition file for the CS
line list \cite{jt615} is given in Table.~\ref{tab:CS_example} and for the
BT2 water line list \cite{jt378} in Table.~\ref{tab:BT2_example}.

\subsubsection{Header}

The first section describes metadata related to internal molecular
data.
$V$ is the version number of the
definition file, it utilises a numerical dating format $YYYYMMDD$ in
order to allow for easy chronological ordering of different versions
by integer comparison. The \texttt{MolKey} is the 27 character InchiKey
\cite{inchikey} used to identify the molecule and $N_{\rm atom}$
describes the number of atoms in the molecule. This allows code to
modify its behaviour based on whether it is dealing with diatomic
molecules $N_{\rm atom}=2$ or polyatomics $N_{\rm atom}>2$.

\begin{longtable}{lrll}
\caption{Definition file format; each entry starts on a new line.} \\
  \label{tab:def_format} \\
\hline 
\multicolumn{1}{c}{Field} & Fortran Format & C Format &   Description \\
\endfirsthead
\multicolumn{3}{c}%
{\tablename\ \thetable\ -- \textit{Continued from previous page}} \\
\hline
Field &  Fortran Format & C Format &   Description \\
\hline
\endhead
\hline \multicolumn{3}{r}{\textit{Continued on next page}} \\
\endfoot
\hline
\endlastfoot
\hline
\multicolumn{4}{l}{\textbf{Header Information}} \\
\texttt{ID} & A10  & \%10s & Always the ASCII string ``EXOMOL.def'' \\
\texttt{IsoFormula} & A27 & \%27s & Isotopologue chemical formula\\
\texttt{Iso-slug} &  A160 & \%160s & Isotopologue identifier, see text for
details\\
\texttt{DSName} &  A10 & \%10s & Isotopologue dataset name\\
$V$ & I8  & \%8d & Version number with format YYYYMMDD\\
\texttt{MolKey} &  A27 & \%27s & Standard inchi key of the molecule \\
$N_{\rm atom}$ & I4  & \%4d & Number of atoms\\
\multicolumn{4}{l}{Atom definition (The following 2 lines occur $N_{\rm atom}$
times)} \\
$I_{\rm atom}$ & I3   & \%3d & Isotope number \\
Atom & A3 & \%3s & Element symbol\\
\multicolumn{4}{l}{\textbf{Isotopologue Information}} \\
$m_{Da} \quad m_{kg}$ & { F12.6,1X,ES14.8}   & {\%12.6f \%14.8e} & Isotopologue mass
in Da and kg \\
$I_{\rm sym}$ & A6 & \%6s & {Molecular symmetry Group (if $N_{\rm atom}=2$ then C or D)}\\
$N_{\rm irrep}$ & I4 & \%4d & Number of irreducible representations\\
\multicolumn{4}{l}{Symmetry definitions (The following 3 lines occur $N_{\rm
irrep}$ times)} \\
$I_{\rm irrep}$ & I3 & \%3d & Irreducible representation ID \\
Symmetry & A6 & \%6s & Irreducible representation label\\
$g_{\rm ns}$ & I3 & \%3d & Nuclear spin degeneracy\\
\multicolumn{4}{l}{\textbf{ExoMol Information}} \\
$T_{\rm max}$ & F8.2 & \%8.2f & Maximum temperature of the line list \\
$N_{\rm broad}$ & I3 & \%3d & No. of pressure broadeners available\\
$D_{\rm avail}$ & I2 & \%2d & Dipole availability (1=Yes 0=No)\\
$N_{\rm cross}$ & I3 & \%3d & No. of cross section files available\\
$N_{\rm kcoef}$ & I3 & \%3d & No. of $k$-coefficient files available\\
\multicolumn{4}{l}{\textbf{States file information}} \\
Life$_{\rm avail}$ & I2 & \%2d & Flag denoting lifetime availability (1=Yes 0=No)\\
$g_{\rm avail}$ & I2 & \%2d & {Flag denoting Land\'e $g$-factor
availability (1=Yes 0=No)}\\
$N_{\rm states}$ & I10 & \%10d & No. of states in \texttt{.states} file\\
$N_{\rm cases}$ & I3 & \%3d & No. of quanta cases\\
\multicolumn{4}{l}{Quanta case definition (The following line occurs $N_{\rm
cases}$ times)}\\
Case Label & A8 & \%8s & Label of the quanta case\\
$N_{\rm quanta}$ & I3 & \%3d & No. of quanta defined\\
\multicolumn{4}{l}{Quanta definition (The following 3 lines occur $N_{\rm
quanta}$ times)} \\
Quanta Label & A3 & \%3s & \\
$F_{\rm Fortran}\quad F_{\rm C}$ & A8,1X,A8 & \%8s \%8s& Fortran and C format for quanta\\
Description & A40 & \%40s & Short description of quanta\\
\multicolumn{4}{l}{\textbf{Trans file definition}} \\
$N_{\rm trans}$ & I15 & \%15d & Total number of transitions\\
$N_{\rm files}$ & I4 & \%4d & Number of transition files \\
$\tilde{v}_{\rm max}$ & F8.2 & \%8.2f & Maximum wavenumber (in \cm)\\
$E_{\rm lower}$ & F8.2 & \%8.2f & Higher energy with complete set of transitions (in $cm^{-1}$)\\
\multicolumn{4}{l}{\textbf{Partition function information}} \\
$T^Q_{\rm max}$ & F8.2 & \%8.2f & Maximum temperature of partition functions \\
$T_{\rm Step}$ & F5.2 & \%5.2f & Temperature step size\\
$C_{\rm avail}$ & I2 & \%2d & Cooling function availability (1=Yes 0=No) \\
\multicolumn{4}{l}{\textbf{Pressure Broadening Information} (default values)} \\
$\gamma_{D}$ & F6.4 & \%6.4f & {\tiny Default value of Lorentzian half-width for
all
lines (in \cm/bar)} \\
$n_{D}$ & F5.3 & \%5.3f & {Default value of the temperature exponent for
all
lines} \\
\multicolumn{4}{l}{\textbf{Pressure Broadening Information} ($N_{\rm broad}>0$,
the following 6 lines occur $N_{\rm broad}$ times)} \\
Broadener Label & A8 & \%8s & Label for particular broadener \\
\multicolumn{2}{l}{Broadener File Name \hfill A20} & \%20s & File name of
particular broadener \\
$J_{\rm max}$ & I7/F7.1 & \%7d/\%7.1f & {\tiny Maximum $J\pp$ for which pressure
broadened
parameters are provided} \\
$\gamma_{L}$ & F6.4 & \%6.4f & {\tiny Value of Lorentzian half-width for lines
with
$J\pp > J_{\rm max}$ (in \cm/bar)}\\
$n_{L}$ & F5.3 & \%5.3f & {\tiny Value of the temperature exponent for lines
with $J\pp > J_{\rm max}$} \\
$N_Q$ & I4 & \%4d & Number of defined quantum number sets \\
\multicolumn{4}{l}{Quantum Sets (The following 3 lines occur $N_{Q}$ times)} \\
Set Code & A2 & \%2s & A code that defines this set of quantum numbers \\
$N_{\rm lines}$ & I6 & \%6d & No. of lines in the broad that contain this code
\\
$N_{\rm Quanta}$ & I4 & \%4d & No. of quantum numbers defined \\
\multicolumn{4}{l}{Quantum Numbers (The following line occurs $N_{\rm Quanta}$
times)} \\
Quantum Label & A4 & \%4s & \\
\hline
\end{longtable}

\subsubsection{Atom Definition}
This section describes each of the atoms in the molecule. The $N_{\rm
  atom}$ field describes how many of them are defined.  
An integer mass number and a string of the
molecule allows a definite description of the isotopologue.

\subsubsection{Isotopologue Information}

This section gives mass and symmetry information about the molecules.
$m$ is the molecular mass given in both Da and kg; the mass is used
for modeling Doppler broadening. Next is the symmetry information of
the molecule. The molecular symmetry group \cite{04BuJexx.method}
 and number of irreducible representations (irreps), $N_{\rm irrep}$,   are
described here, after which $N_{\rm irrep}$ lines defining each irrep
in that group are given. The $I_{\rm sym}$ field gives the
ID of the symmetry used in the \texttt{.states} file. However, for
linear molecules including diatomics the symmetry group label will
only take values of \textit{C} or \textit{D} corresponding to
$C_{\infty v}$ or $D_{\infty h}$ respectively and $N_{\rm sym}$ is set
to 0.  $g_{\rm ns}$ corresponding to nuclear spin degenracy will be
provided for \textit{C}.  For \textit{D}, $g_{\rm even}$ and $g_{\rm
  odd}$ are provided; these correspond to the nuclear-spin degeneracy
factor needed for when the rovibronic wavefunction (i.e.  the
wavefunction without nuclear spin) is respectively even or odd with
respect to interchange of identical atoms.  Otherwise we follow
symmetry labels and conventions as set out by
Bunker and Jensen \cite{04BuJexx.method}.
In this case $g_{\rm ns}$ is be provided for each
irreducible representation.

\subsubsection{ExoMol Information}
This section relates to the availability of certain aspects of the
line list. First and foremost it describes the maximum temperature for
which the line list is applicable ($T_{\rm max}$).  Secondly, for
molecules with pressure broadening parameters, it describes the number
of broadeners ($N_{\rm broad}$) available; this is used later on to
determine what additional structures should be parsed in the file.
Finally, there is a boolean field which describes whether we provide
dipoles for the particular molecule.  

\subsubsection{State file information}

This section is the most important as it describes how the
\texttt{.states} file format is structured. The first two fields
describe whether lifetime information and Land\'e $g$-factors are
available. These will be placed after the initial `standard' fields in
the states file.  Then there is the $N_{\rm states}$ field which gives
the number of states in the file. Finally the number of quanta cases,
$N_{\rm cases}$. A quanta case describes what the following quantum
numbers represent. For example, commonly for polyatomics the ExoMol
\texttt{.states} file provides two representations of quantum numbers;
the standard normal mode quantum numbers and the local mode `TROVE'
quantum numbers. For this example $N_{\rm cases}=2$; see Down \ea\
\cite{jt546} for  instance; a recent review of local modes has
been given by Jensen \cite{12Jensen.cluster}. Each case is given a label and must
describe the number of quanta defined $N_{\rm quanta}$, after which
$N_{\rm quanta}$ lines will follow, each describing a particular
quantum number. Quantum numbers are described by a string label
comprising a string format identifier and a short, human-readable
description. The string format identifier utilises the FORTRAN string
formatting to allow for the construction of formats more easily for
FORTRAN which has stricter string handling than C/C++. An overview of
the new \texttt{.states} format is given in Section \ref{ss:states}.

\subsubsection{Transition and partition information}
This section simply describes the availability of \texttt{.trans} and
\texttt{.pf} files.  In particular, for the partition function file
(\texttt{.pf}) it describes the maximum temperature provided ($T^Q_{\rm
  max}$), the step size of the temperature ($T_{\rm Step}$) and whether
a cooling function is also present.

\subsubsection{Pressure broadening information}

Finally, the pressure broadening information is provided here. If
$N_{\rm broad}=0$ this simply contains default values of Lorentzian
half-width $\gamma_{L}$ and temperature exponent $n_{L}$. If $N_{\rm
  broad}>0$ this section has the following structure: the broadener is
defined first, immediately followed by the associated file name,
$J_{\rm max}$, and the asymptotic values of the Lorentzian half-width
$\gamma_{D}$ and temperature exponent $n_{D}$ for the $a0$ quantum
number set  (see Section 5.6). $N_Q$ quantum number sets are explicitly defined by a
label (e.g. $a0$) that corresponds to codes in the broadener file, the
number of lines in the broadener file with that code and the number of
quantum numbers in that set. After this each additional quantum number is also
described;
the total angular momentum quantum number of the lower state $J"$ is
compulsory.
 The quantum numbers relate to the quanta defined previously
in the states section followed by either a prime or double prime.
Further details relating to pressure broadening are given in Section
\ref{ss:broad}.

The final field of the definition file is used for keywords to facilitate
the search of the database by web search engines.

\SaveVerb{term}|12C-32S__JnK.def|
\begin{longtable}{ll}
\hline
\verb!  EXOMOL.def                            !&\verb!#   ID    !\\
\verb!  (12C)(32S)                            !&\verb!#   IsoFormula      !\\
\verb!  12C-32S                               !&\verb!#   Iso-slug       !\\
\verb!  JnK                                   !&\verb!#   Isotopologue data set name        !\\
\verb!  20160217                              !&\verb!#   Version number with format YYYYMMDD  !\\
\verb!  DXHPZXWIPWDXHJ-UHFFFAOYSA-N           !&\verb!#   Inchi key of molecule   !\\
\verb!  2                                     !&\verb!#   Number of atoms    !\\
\verb!  12                                    !&\verb!#   Isotope number 1       !\\
\verb!  C                                     !&\verb!#   Element symbol 1      !\\
\verb!  32                                    !&\verb!#   Isotope number 2      !\\
\verb!  S                                     !&\verb!#   Element symbol 2     !\\
\verb!     43.972071 7.30173406e-26           !&\verb!#   Isotopologue mass (Da) and (kg)       !\\
\verb!  C                                     !&\verb!#   Symmetry group        !\\
\verb!  2                                     !&\verb!#   Number irreducible representations     !\\
\verb!  1                                     !&\verb!#   Irreducible representation ID        !\\
\verb!  Sigma+                                !&\verb!#   Irreducible representation label     !\\
\verb!  1                                     !&\verb!#   Nuclear spin degeneracy     !\\
\verb!  2                                     !&\verb!#   Irreducible representation ID     !\\
\verb!  Sigma-                                !&\verb!#   Irreducible representation label     !\\
\verb!  1                                     !&\verb!#   Nuclear spin degeneracy    !\\
\verb!   3000                                 !&\verb!#   Maximum temperature of linelist      !\\
\verb!  2                                     !&\verb!#   No. of pressure broadeners available   !\\
\verb!  0                                     !&\verb!#   Dipole availability (1=yes, 0=no)      !\\
\verb!  0                                     !&\verb!#   No. of cross section files available !\\
\verb!  0                                     !&\verb!#   No. of k-coefficient files available   !\\
\verb!  0                                     !&\verb!#   Lifetime availability (1=yes, 0=no)   !\\
\verb!  0                                     !&\verb!#   Lande g-factor availability (1=yes, 0=no) !\\
\verb!  11497                                 !&\verb!#   No. of states in .states file     !\\
\verb!  1                                     !&\verb!#   No. of quanta cases          !\\
\verb!  dcs                                   !&\verb!#   Quantum case label                                                      !\\
\verb!  1                                     !&\verb!#   No. of quanta defined                                                   !\\
\verb!  v                                     !&\verb!#   Quantum label 1                                                        !\\
\verb!  I4 %4d                                !&\verb!#   Format quantum label 1                                                  !\\
\verb!  State vibrational quantum number      !&\verb!#   Description quantum label 1                                             !\\
\verb!  199045                                !&\verb!#   Total number of transitions                                             !\\
\verb!  1                                     !&\verb!#   No. of transition files                                                 !\\
\verb!  10996.09                              !&\verb!#   Maximum wavenumber (in cm-1)                                            !\\
\verb!  30000.00                           !&\verb!#   Higher energy with complete set of transitions (in cm-1)     !\\
\verb!   3000.0                               !&\verb!#   Maximum temperature of partition function                               !\\
\verb!   1.00                                 !&\verb!#   Temperature step size                                                !\\
\verb!  0                                     !&\verb!#   Cooling function availability (1=yes, 0=no)                             !\\
\verb!  0.0700                                !&\verb!#   Default value of Lorentzian half-width for all lines (in cm-1/bar)      !\\
\verb!  0.500                                 !&\verb!#   Default value of temperature exponent for all lines                     !\\
\verb!  air                                   !&\verb!#   Label for a particular broadener                                        !\\
\verb!  12C-32S__air.broad                    !&\verb!#   Filename of particular broadener                                        !\\
\verb!  46                                    !&\verb!#   Maximum J for which pressure broadening parameters provided             !\\
\verb!  0.0698                                !&\verb!#   Value of Lorentzian half-width for J" > Jmax                            !\\
\verb!  0.750                                 !&\verb!#   Value of temperature exponent for lines with J" > Jmax                  !\\
\verb!  1                                     !&\verb!#   Number of defined quantum number sets                                   !\\
\verb!  a0                                    !&\verb!#   A code that defines this set of quantum numbers                         !\\
\verb!  47                                    !&\verb!#   No. of lines in the broad that contain this code                        !\\
\verb!  0                                     !&\verb!#   No. of quantum numbers defined                                          !\\
\verb!  self                                  !&\verb!#   Label for a particular broadener                                        !\\
\verb!  12C-32S__self.broad                   !&\verb!#   Filename of particular broadener                                        !\\
\verb!  46                                    !&\verb!#   Maximum J for which pressure broadening parameters provided             !\\
\verb!  0.0620                                !&\verb!#   Value of Lorentzian half-width for J" > Jmax                            !\\
\verb!  0.500                                 !&\verb!#   Value of temperature exponent for lines with J" > Jmax                  !\\
\verb!  1                                     !&\verb!#   Number of defined quantum number sets                                   !\\
\verb!  a0                                    !&\verb!#   A code that defines this set of quantum numbers                         !\\
\verb!  47                                    !&\verb!#   No. of lines in the broad that contain this code                        !\\
\verb!  0                                     !&\verb!#   No. of quantum numbers defined                                          !\\
\hline
\caption{File \protect\UseVerb{term}:  the definition file for $^{12}$C$^{32}$S.}\label{tab:CS_example}
\end{longtable}

\SaveVerb{term}|BT2.def|
\begin{longtable}{ll}
\hline
\verb!  EXOMOL.def                              !&\verb!   #  ID    !\\
\verb!  (1H)2(16O)                              !&\verb!   #  IsoFormula    !\\
\verb!  1H2-16O                                 !&\verb!   #  Iso-slug         !\\
\verb!  BT2                                     !&\verb!   #  Isotopologue   !\\
\verb!  20160220                                !&\verb!   #  Version number with format YYYYMMDD      !\\
\verb!  XLYOFNOQVPJJNP-UHFFFAOYSA-N             !&\verb!   #  Inchi key of molecule                                                           !\\
\verb!  3                                       !&\verb!   #  Number of atoms                                                                 !\\
\verb!  1                                       !&\verb!   #  Isotope number 1                                                                !\\
\verb!  H                                       !&\verb!   #  Element symbol 1                                                                !\\
\verb!  16                                      !&\verb!   #  Isotope number 2                                                                !\\
\verb!  O                                       !&\verb!   #  Element symbol 2                                                                !\\
\verb!     18.010565 2.99072463e-26             !&\verb!   #  Isotopologue mass (Da) and (kg)                                                 !\\
\verb!  C2v                                     !&\verb!   #  Symmetry group                                                                  !\\
\verb!  4                                       !&\verb!   #  Number of irreducible representations                                           !\\
\verb!  1                                       !&\verb!   #  Irreducible representation ID                                                   !\\
\verb!  A1                                      !&\verb!   #  Irreducible representation label                                                !\\
\verb!  1                                       !&\verb!   #  Nuclear spin degeneracy                                                         !\\
\verb!  2                                       !&\verb!   #  Irreducible representation ID                                                   !\\
\verb!  A2                                      !&\verb!   #  Irreducible representation label                                                !\\
\verb!  1                                       !&\verb!   #  Nuclear spin degeneracy                                                         !\\
\verb!  3                                       !&\verb!   #  Irreducible representation ID                                                   !\\
\verb!  B1                                      !&\verb!   #  Irreducible representation label                                                !\\
\verb!  3                                       !&\verb!   #  Nuclear spin degeneracy                                                         !\\
\verb!  4                                       !&\verb!   #  Irreducible representation ID                                                   !\\
\verb!  B2                                      !&\verb!   #  Irreducible representation label                                                !\\
\verb!  3                                       !&\verb!   #  Nuclear spin degeneracy                                                         !\\
\verb!   3000.0                                 !&\verb!   #  Maximum temperature of linelist                                                 !\\
\verb!  0                                       !&\verb!   #  No. of pressure broadeners available                                            !\\
\verb!  0                                       !&\verb!   #  Dipole availability (1=yes, 0=no)                                               !\\
\verb!  0                                       !&\verb!   #  No. of cross section files available                                    !\\
\verb!  0                                       !&\verb!   #  No. of k-coefficient files available                                    !\\
\verb!  0                                       !&\verb!   #  Lifetime availability (1=yes, 0=no)                                             !\\
\verb!  0                                       !&\verb!   #  Lande g-factor availability (1=yes, 0=no)                                       !\\

\verb!  221097                                  !&\verb!   #  No. of states in .states file                                                   !\\
\verb!  1                                       !&\verb!   #  No. of quanta cases                                                             !\\
\verb!  nltcs                                   !&\verb!   #  Quantum case label                                                              !\\
\verb!  10                                      !&\verb!   #  No. of quanta defined                                                           !\\
\verb!  +/-                                     !&\verb!   #  Quantum label 1                                                                 !\\
\verb!  A1 %1s                                  !&\verb!   #  Format quantum label 1                                                          !\\
\verb!  Total parity                            !&\verb!   #  Description quantum label 1                                                     !\\
\verb!  Gamma                                   !&\verb!   #  Quantum label 3                                                                 !\\
\verb!  I2 %2d                                  !&\verb!   #  Format Quantum label 2                                                          !\\
\verb!  Symmetry block number (1)               !&\verb!   #  Description Quantum label 2                                                     !\\
\verb!  N                                       !&\verb!   #  Quantum label 3                                                                 !\\
\verb!  I10 %10d                                !&\verb!   #  Format Quantum label 3                                                          !\\
\verb!  Reference within the symmetry block     !&\verb!   #  Description Quantum label 4                                                     !\\
\verb!  nucspin                                 !&\verb!   #  Quantum label 4                                                                 !\\
\verb!  A1 %1s                                  !&\verb!   #  Format Quantum label 4                                                          !\\
\verb!  Nuclear spin isomer label(2)            !&\verb!   #  Description Quantum label 5                                                     !\\
\verb!  Gamma_rve                               !&\verb!   #  Quantum label 5                                                                 !\\
\verb!  A2 %2s                                  !&\verb!   #  Format Quantum label 5                                                          !\\
\verb!  Rovibrational symmetry label            !&\verb!   #  Description Quantum label 6                                                     !\\
\verb!  v1                                      !&\verb!   #  Quantum label 6                                                                !\\
\verb!  I2 %2d                                  !&\verb!   #  Format Quantum label 6                                                         !\\
\verb!  v1 symmetric stretch quantum number(3   !&\verb!   #  Description Quantum label 7                                                    !\\
\verb!  v2                                      !&\verb!   #  Quantum label 7                                                                !\\
\verb!  I2 %2d                                  !&\verb!   #  Format Quantum label 7                                                         !\\
\verb!  v2 bend quantum number                  !&\verb!   #  Description Quantum label 8                                                    !\\
\verb!  v3                                      !&\verb!   #  Quantum label 8                                                                !\\
\verb!  I2 %2d                                  !&\verb!   #  Format Quantum label 8                                                         !\\
\verb!  v2 asymmetric stretch quantum number    !&\verb!   #  Description Quantum label 9                                                    !\\
\verb!  Ka                                      !&\verb!   #  Quantum label 9                                                                !\\
\verb!  I2 %2d                                  !&\verb!   #  Format Quantum label 9                                                         !\\
\verb!  Ka rotational quantum number            !&\verb!   #  Description Quantum label 10                                                    !\\
\verb!  Kc                                      !&\verb!   #  Quantum label 10                                                                !\\
\verb!  I2 %2d                                  !&\verb!   #  Format Quantum label 10                                                         !\\
\verb!  Kc rotational quantum number            !&\verb!   #  Description quantum label 10                                                    !\\
\verb!  505806255                               !&\verb!   #  Total number of transitions                                                     !\\
\verb!  16                                      !&\verb!   #  No. of transition files                                                         !\\
\verb!  29971.78                                !&\verb!   #  Maximum wavenumber (in cm-1)                                                    !\\
\verb!     20000.00                                !&\verb!   #  Higher energy with complete set of transitions (in cm-1)                 !\\
\verb!   3000.0                                 !&\verb!   #  Maximum temperature of partition function                                       !\\
\verb!   1.0                                    !&\verb!   #  Step size of temperature                                                        !\\
\verb!  1                                       !&\verb!   #  Cooling function availability (1=yes, 0=no)                                     !\\
\verb!  0.0700                                  !&\verb!   #  Default value of Lorentzian half-width for all lines (in cm-1/bar)              !\\
\verb!  0.500                                   !&\verb!   #  Default value of temperature exponent for all lines                             !\\
\hline
\caption{File \protect\UseVerb{term}:  the definition file for
H$_2$$^{16}$O.}\label{tab:BT2_example}
\end{longtable}

\subsection{States file}\label{ss:states}

The basic ExoMol data structure is a
comprehensive list of the molecular states involved. This file
contains a numbered list of energy terms, in cm$^{-1}$ with the
zero defined by the lowest level, immediately followed by the total
statistical weight $g_\mathrm{tot} = (2J+1) \times g_{\rm ns}$ in the HITRAN
convention
\cite{HITRAN-A} which retains the full nuclear-spin degeneracy of all
atoms.
These three columns are sufficient, in conjunction
with the transitions file, to model absorption and/or emission
intensities.  The degeneracy is followed by the
total angular momentum quantum number $J$
(integer or half-integer). The next
columns are used for the state lifetimes (s) and/or Land\'e
$g$-factors with availability determined in the definition file.
This is followed by good (rigorous) quantum number(s) such as
the state symmetry,
parity or a total symmetry (irreducible representation).
Then follows any approximate quantum numbers that are specified. There is some
flexibility over this portion of the file which is defined on a
case-by-case basis, see Ref.~\cite{VAMDCcasebycase} for example.

For polyatomic molecules the approximate quantum numbers include, for example,
the vibrational normal mode $v_i$, local mode $n_i$ quantum numbers,
rotational quantum numbers $K$, $K_a$, $K_c$, polyad numbers,
vibrational and rotational symmetries etc. For the diatomic molecules
these are the spin components $F_i$, $\Sigma$ or $\Omega$, electronic
term designations, e.g $^{3}{}\Pi_{g}$, vibrational quantum numbers
$v$, rotational angular momentum $N$ (Hund's case~(b)), projection of
the electronic angular momentum, $\Lambda$, etc. The spin components quantum
numbers can be integral, half-integral or even strings F1, F2,
$\ldots$ If the theoretical energy is replaced by an experimental
counterpart, the former is usually kept as an extra field.

Regardless of molecule, the structure of any ExoMol state file can be
determined from its complimentary definition file. The structure in
order is:
\begin{enumerate}
 \item Mandatory fields
 \item Lifetime if Life$_{\rm avail}=1$
 \item Land\'e $g$-factor if $g_{\rm avail}=1$
 \item Quanta Cases in order of definition
\end{enumerate}

The detailed specification of the mandatory section of the states file
is given in Table~\ref{Tab:states}. Using the state file information
from Table~\ref{tab:CS_example}, the mandatory definition in
the FORTRAN string format can easily be
programmatically generated:
\begin{verbatim}
"(I12,1x,F12.6,1x,I6,1x,I7,1x,ES12.4,1x,A1,1x,A1,1x,A9,1x,I2,1x,I2,1x,I2,1x,I2)"
\end{verbatim}
This format corresponds to the sample state file given in
Table~\ref{tab:levels}.

Note that for lifetimes  it is necessary to deal with states
with infinite radiative lifetimes (e.g. the ground state) and for which
lifetimes are not available: for high-lying states not all downward
transitions are considered so it is not possible to compute the
lifetime. Here, and elsewhere in the database,
infinity is specified by the string \lq INF' and unknown
numbers by \lq NaN'.

\begin{table}
\caption{Specification of the mandatory part of the states file and
extra data options.}
\label{Tab:states}
\begin{tabular}{llll} \hline
Field & Fortran Format & C Format & Description\\ \hline
$i$ & \texttt{I12} & \texttt{\%12d} & State ID\\
$E$ & \texttt{F12.6} & \texttt{\%12.6f} & State energy in $\mathrm{cm^{-1}}$\\
$g_\mathrm{tot}$ & \texttt{I6} & \texttt{\%6d} & State degeneracy\\
$J$ & \texttt{I7/F7.1} & \texttt{\%7d/\%7.1f} & $J$-quantum number
(integer/half-integer)\\
($\tau$) & \texttt{ES12.4} & \texttt{\%12.4e} & Lifetime in s (optional)\\
($g$) & \texttt{F10.6} & \texttt{\%10.6f} & Land\'e $g$-factor  (optional)\\
\hline
\end{tabular}

\noindent
 {\flushleft
ID: state identifier: a non-negative integer index, starting at 1\\
$J$:  total angular momentum quantum,  excluding nuclear spin \\

\noindent
Fortran format, $J$ integer:
\texttt{(I12,1x,F12.6,1x,I6,I7,1x,ES12.4,1x,F10.6)}\\
or $J$ half-integer:  \texttt{(I12,1x,F12.6,1x,I6,F7.1,1x,ES12.4,1x,F10.6)}\\
}

\end{table}

\SaveVerb{term}|40Ca-16O__VBATHY.state|
\begin{table}
\caption{Extract from the state file for  $^{40}$Ca$^{16}$O,
\protect\UseVerb{term}.}
\label{tab:levels}
\begin{center}
\footnotesize
\tabcolsep=5pt
\begin{tabular}{rrrrccrrlrrrr c rrr}
\hline
     $i$ & \multicolumn{1}{c}{$\tilde{E}$} &  $g$    & $J$  & $\tau$ & $g$
&\multicolumn{1}{c}{$+/-$} &  \multicolumn{1}{c}{$e/f$} & State & $v$ &
$\Lambda$& $\Sigma$ & $\Omega$& \\
\hline
     4051  &     8242.235601  &     15  &      7  &      6.5090E-01  &
0.106733  &     +    &     f   &\verb!  a3Pi         !&      0  &       1  &
  1  &       2     \\
     4052  &     8294.404892  &     15  &      7  &      2.1060E-03  &
0.018230  &     +    &     f   &\verb!  a3Pi         !&      0  &       1  &
  0  &       1     \\
     4053  &     8369.059680  &     15  &      7  &      4.5940E-01  &
0.000045  &     +    &     f   &\verb!  a3Pi         !&      0  &       1  &
 -1  &       0     \\
     4054  &     8627.261523  &     15  &      7  &      6.4790E-05  &
0.017893  &     +    &     f   &\verb!  Ap1Pi        !&      0  &       1  &
  0  &       1     \\
     4055  &     8781.692232  &     15  &      7  &      1.8810E-01  &
0.106728  &     +    &     f   &\verb!  a3Pi         !&      1  &       1  &
  1  &       2     \\
     4056  &     8833.629438  &     15  &      7  &      1.4660E-03  &
0.018239  &     +    &     f   &\verb!  a3Pi         !&      1  &       1  &
  0  &       1     \\
     4057  &     8907.535275  &     15  &      7  &      1.5950E-01  &
0.000048  &     +    &     f   &\verb!  a3Pi         !&      1  &       1  &
 -1  &       0     \\
     4058  &     9167.417952  &     15  &      7  &      4.4060E-05  &
0.017996  &     +    &     f   &\verb!  Ap1Pi        !&      1  &       1  &
  0  &       1     \\
     4059  &     9314.686162  &     15  &      7  &      1.0190E-01  &
0.106678  &     +    &     f   &\verb!  a3Pi         !&      2  &       1  &
  1  &       2     \\
     4060  &     9365.515048  &     15  &      7  &      1.1390E-03  &
0.018390  &     +    &     f   &\verb!  a3Pi         !&      2  &       1  &
  0  &       1     \\
     4061  &     9434.128287  &     15  &      7  &      7.8210E-02  &
0.000113  &     +    &     f   &\verb!  a3Pi         !&      2  &       1  &
 -1  &       0     \\
     4062  &     9533.168336  &     15  &      7  &      2.1600E-03  &
0.029567  &     +    &     f   &\verb!  b3Sigma+     !&      0  &       0  &
  1  &       1     \\
     4063  &     9545.896549  &     15  &      7  &      1.1140E-02  &
0.005414  &     +    &     f   &\verb!  b3Sigma+     !&      0  &       0  &
  0  &       0     \\
     4064  &     9709.239641  &     15  &      7  &      3.3420E-05  &
0.018256  &     +    &     f   &\verb!  Ap1Pi        !&      2  &       1  &
  0  &       1     \\
     4065  &     9841.636252  &     15  &      7  &      8.9050E-02  &
0.106743  &     +    &     f   &\verb!  a3Pi         !&      3  &       1  &
  1  &       2     \\
     4066  &     9893.716709  &     15  &      7  &      1.0390E-03  &
0.018284  &     +    &     f   &\verb!  a3Pi         !&      3  &       1  &
  0  &       1     \\
     4067  &     9966.036778  &     15  &      7  &      6.8230E-02  &
0.000061  &     +    &     f   &\verb!  a3Pi         !&      3  &       1  &
 -1  &       0     \\
     4068  &    10103.239512  &     15  &      7  &      1.1060E-03  &
0.026224  &     +    &     f   &\verb!  b3Sigma+     !&      1  &       0  &
  1  &       1     \\
     4069  &    10115.858820  &     15  &      7  &      2.8870E-03  &
0.008900  &     +    &     f   &\verb!  b3Sigma+     !&      1  &       0  &
  0  &       0     \\
     4070  &    10240.534926  &     15  &      7  &      2.7290E-05  &
0.018317  &     +    &     f   &\verb!  Ap1Pi        !&      3  &       1  &
  0  &       1     \\
\hline
\end{tabular}
\end{center}

\noindent
{\flushleft
$i$:   State counting number.     \\
$\tilde{E}$: State energy in \cm. \\
$g$: State degeneracy.            \\
$J$: Total angular momentum.\\
$\tau$: Lifetime in s.\\
$g$:   Land\'e $g$-factor   \\
$+/-$:   Total parity. \\
$e/f$:   rotationless-parity. \\
$v$:   State vibrational quantum number. \\
$\Lambda$:  Projection of the electronic angular momentum. \\
$\Sigma$:   Projection of the electronic spin. \\
$\Omega$:   $\Omega=\Lambda+\Sigma$, projection of the total angular momentum.\\
}

\end{table}

\subsection{Transitions file}

The transitions file has a simple structure, see
Table~\ref{tab:trans}. Two pointers, $i$ and $f$ point to rows in
the \texttt{.states} file to identify the upper and lower states involved plus
the full information characterising these states. The $A$ gives the
Einstein $A$ coefficient for this transition. This file can be very
large but, optionally, there is a fourth column $\tilde{\nu}$ which
gives the transition wavenumber. Otherwise this must be computed from
the states file as the difference of the upper and lower energy
levels.  A sample transitions file is given in
Table~\ref{tab:transeg}.

Many of the molecules in the database are characterised by a very large
number of transitions.  To make the use of \texttt{.trans} tractable
these files are often split into wavenumber regions, as discussed
above. The \texttt{.trans} files are also compressed using
\texttt{.bz2} format; a utility is provided for reading this format
without needing to uncompress the file.

\begin{table}
\caption{Specification of the transitions file.}
\label{tab:trans}
\begin{tabular}{llll} \hline
Field & Fortran Format & C Format & Description\\ \hline
$i$ & \texttt{I12} & \texttt{\%12d} & Upper state ID\\
$f$ & \texttt{I12} & \texttt{\%12d} & Lower state ID\\
$A$ & \texttt{ES10.4} & \texttt{\%10.4e} & Einstein $A$ coefficient in
$\mathrm{s^{-1}}$ \\
$\tilde{\nu}_{fi}$&\texttt{F15.6} & \texttt{\%15.6f} & Transition wavenumber in
cm$^{-1}$ (optional). \\
 \hline
\end{tabular}

\noindent
Fortran format: \texttt{(I12,1x,I12,1x,ES10.4,1x,ES15.6)}\\
\end{table}

\SaveVerb{term}|40Ca-16O__VBATHY.trans|
\begin{table}
\caption{  Extract from the transitions file for $^{40}$Ca$^{16}$O,
\protect\UseVerb{term}.}
\label{tab:transeg}
\begin{center}
\begin{tabular}{rrcc}
\hline\hline
$i$     &  $f$          &               $A_{if}$ &  $\tilde{\nu}_{if}$\\
\hline
       10571    &        10884   &    9.5518E-06       &        120.241863\\
       21053    &        21375   &    1.9515E-05       &        120.242886\\
        8726    &         9672   &    1.8658E-04       &        120.243522\\
       11655    &        11950   &    5.0065E-06       &        120.243733\\
       93209    &        93967   &    5.7055E-03       &        120.244192\\
        2228    &         3175   &    7.3226E-07       &        120.244564\\
       46727    &        46432   &    1.0599E-04       &        120.244658\\
       44436    &        44774   &    1.4626E-04       &        120.245583\\
       29037    &        28723   &    1.8052E-04       &        120.245669\\
        4458    &         4805   &    1.0431E-08       &        120.246396\\
       69313    &        68434   &    5.0531E-06       &        120.248178\\
       22640    &        22985   &    1.1281E-07       &        120.248891\\
       57027    &        56721   &    7.1064E-06       &        120.250180\\
   \hline
\end{tabular}

\noindent
$i$: Upper  state counting number;\\
$f$:  Lower  state counting number; $A_{if}$:  Einstein $A$\\
coefficient in s$^{-1}$; $\tilde{\nu}_{if}$: transiton wavenumber in \cm.
\end{center}
\end{table}

\subsection{Pressure broadening file}\label{ss:broad}

\begin{table}
\caption{Specification of the mandatory part of the pressure broadening
parameters file.}
\label{tab:broad_format} \footnotesize
\begin{center}
\begin{tabular}{llll}
\hline
Field & Fortran Format & C format & Description \\
\hline
code & A2 & \%2s & Code identifying quantum number set following $J\pp$ \\

$\gamma_{\textrm{ref}}$ & F6.4 & \%6.4f & Lorentzian half-width at reference 
temperature and pressure in \cm/bar \\
$n$ & F6.3 & \%6.3f & Temperature exponent \\
$J\pp$ & I7/F7.1 & \%7d/\%7.1f & Lower $J$-quantum number integer/half-integer
\\
\hline
\end{tabular}
\end{center}

\noindent
Fortran format, $J$ integer: \texttt{(A2,1x,F6.4,1x,F6.3,1x,I7)}\\
or $J$ half-integer: \texttt{(A2,1x,F6.4,1x,F6.3,1x,F7.1)}\\
\end{table}

Like the \texttt{.states} file, the first four fields of the \texttt{.broad}
file are mandatory for all records, see
Table~\ref{tab:broad_format}. This includes one quantum number,
$J\pp$, which is always known and hence guarantees that at least
a semi-empirical Lorentzian half-width ($\gamma_{\textrm{ref}}$) and
temperature dependence, represented by exponent $n$
(see Eq.~\ref{eq:TP_dep}), can be generated for every
molecular line. Additional upper and lower state labels,
which give molecule and quantum number dependent behavior,
follow the compulsory fields. These  are given in the
definitions file along with  the maximum $J\pp$ for which approximate
parameters have been generated, $J\pp_{\rm max}$, and the respective
values of Lorentzian half-width and temperature exponent,
$\gamma_{\rm max}$ and $n_{\rm max}$.  When $J\pp > J\pp_{\rm max}$,
$\gamma_{\rm max}$ and $n_{\rm max}$ should be used for the
Lorentzian half-width and temperature exponent respectively.

The \texttt{.broad} file has a hierarchical structure; values of
$\gamma_{\textrm{ref}}$ and $n$ with full quantum assignments are
presented first, followed by values with partial quantum
assignments, then finally values with $J\pp$ dependence only.
This represents the preferential order in which the values of
$\gamma_{\textrm{ref}}$ and $n$ should be used. This hierarchical
structure is demonstrated in Table~\ref{tab:broad_H2O_example}
where a portion of the H$_2$O-H$_2$ \texttt{.broad} file is presented.

Once a suitable $\gamma_{\rm ref}$ and $n$ have been identified,
the Lorentzian half-width of a spectral line at temperature $T$
and pressure $P$ can be calculated as:
\begin{equation}
\label{eq:TP_dep}
\gamma(T) = \gamma_{\textrm{ref}} \times
\left(\frac{T_{\textrm{ref}}}{T}\right)^{n} \times
\left(\frac{P}{P_{\textrm{ref}}}\right).
\end{equation}
Here, $T_{\textrm{ref}}$ = 296 K and $P_{\textrm{ref}}$ = 1 bar.
Note that to convert from \cm/bar used by ExoMol to \cm/atm used by Hitran
requires $\gamma$ to be mutiplied by 1.01325.

A separate \texttt{.broad} file is supplied for each molecule-broadener
system and the broadener is specified in the file name. Example\texttt{.broad} files for CS-Air and H$_2$O-H$_2$ are given in
Table~\ref{tab:broad_CS_example} and Table~\ref{tab:broad_H2O_example}.
For particular molecule-broadener systems where no pressure broadening
information is available, the default values of $\gamma_{\textrm{ref}}$
and $n$ given in the isotopologue's definition file may be used.

\SaveVerb{term}|12C-32S__air.broad|
\begin{table}
\caption{File \protect\UseVerb{term}: Air \texttt{.broad} file for
$^{12}$C$^{32}$S:
portion of the file (upper part); field specification (lower part).}
\label{tab:broad_CS_example} \footnotesize
\begin{center}
\begin{tabular}{llll}
\hline
a0 & 0.0860 & 0.096 & \hsn0 \\
a0 & 0.0850 & 0.093 & \hsn1 \\
a0 & 0.0840 & 0.091 & \hsn2 \\
a0 & 0.0840 & 0.089 & \hsn3 \\
a0 & 0.0830 & 0.087 & \hsn4 \\
... & & & \\
a0 & 0.0720 & 0.067 & \hsx35 \\
a0 & 0.0720 & 0.066 & \hsx36 \\
... & & & \\
\end{tabular}
\begin{tabular}{llll}
\hline
Field & Fortran Format & C format & Description \\
\hline
code & A2 & $\%$2s & Code identifying quantum number set following $J\pp$* \\
$\gamma_{\textrm{ref}}$ & F6.4 & \%6.4f & Lorentzian half-width at reference
temperature and pressure in \cm/bar \\
$n$ & F5.3 & \%5.3f & Temperature exponent \\
$J\pp$ & I7/F7.1 & $\%$7d & Lower $J$-quantum number \\
\hline
\end{tabular}
\end{center}
\noindent
*Code definition:
a0 = none
\end{table}                                     

\SaveVerb{term}|1H2-16O__H2.broad|
\begin{table}
\caption{\protect\UseVerb{term}: H$_{2}$O - H$_{2}$ broad file: portion of the
file (upper part); field specification (lower part).}
\label{tab:broad_H2O_example} \footnotesize
\begin{center}
\begin{tabular}{lllllllllllllll}
\hline
... & & & & & & & & & & & & & & \\
b2 & 0.0356 & 0.300 & \hsn9 & \hsn8 & \hto1 & \hto9 & \hto0 & \hto8 & \hto0 &
\hto0 & \hto0 & \hto0 & \hto1 & \hto0 \\
b2 & 0.0522 & 0.300 & \hsn9 & \hsn8 & \hto1 & \hto9 & \hto0 & \hto8 & \hto0 &
\hto0 & \hto0 & \hto1 & \hto0 & \hto0 \\
b2 & 0.0521 & 0.300 & \hsn7 & \hsn8 & \hto1 & \hto6 & \hto1 & \hto7 & \hto0 &
\hto0 & \hto0 & \hto0 & \hto0 & \hto1 \\
... & & & & & & & & & & & & & & \\
a5 & 0.0600 & 0.546 & \hsn3 & \hsn4 & \hto0 & \hto3 & \hto1 & \hto4 &&&&&& \\
a5 & 0.0618 & 0.551 & \hsn3 & \hsn4 & \hto2 & \hto1 & \hto1 & \hto4 &&&&&& \\
a5 & 0.0569 & 0.525 & \hsn2 & \hsn3 & \hto2 & \hto1 & \hto3 & \hto0 &&&&&& \\
... & & & & & & & & & & & & & & \\
a1 & 0.0301 & 0.268 & \hsx14 & \hsx15 &&&&&&&&&& \\
a1 & 0.0291 & 0.230 & \hsx15 & \hsx16 &&&&&&&&&& \\
a1 & 0.0282 & 0.218 & \hsx16 & \hsx17 &&&&&&&&&& \\
... & & & & & & & & & & & & & & \\
a0 & 0.0242 & 0.165 & \hsx24 &&&&&&&&&&& \\
a0 & 0.0239 & 0.160 & \hsx25 &&&&&&&&&&& \\
a0 & 0.0236 & 0.150 & \hsx26 &&&&&&&&&&& \\
... & & & & & & & & & & & & & & \\
\end{tabular}
\begin{tabular}{llll}
\hline
Field & Fortran Format & C format & Description \\
\hline
code & A2 & $\%$2s & Code identifying quantum number set following $J\pp$* \\
$\gamma_{\textrm{ref}}$ & F2.4 & \%2.4f & Lorentzian half-width at reference
temperature and pressure in \cm/bar \\
$n$ & F2.3 & \%2.3f & Temperature exponent \\
$J\pp$ & I7 & $\%$7d & Lower $J$-quantum number \\
$J\p$ & I7 & $\%$7d & Upper $J$-quantum number \\
$K_a\pp$ & I2 & $\%$2d & Lower rotational quantum number \\
$K_c\pp$ & I2 & $\%$2d & Lower rotational quantum number \\
$K_a\p$ & I2 & $\%$2d & Upper rotational quantum number \\
$K_c\p$ & I2 & $\%$2d & Upper rotational quantum number \\
$v_1\pp$ & I2 & $\%$2d & Lower vibrational quantum number \\
$v_2\pp$ & I2 & $\%$2d & Lower vibrational quantum number \\
$v_3\pp$ & I2 & $\%$2d & Lower vibrational quantum number \\
$v_1\p$ & I2 & $\%$2d & Upper vibrational quantum number \\
$v_2\p$ & I2 & $\%$2d & Upper vibrational quantum number \\
$v_3\p$ & I2 & $\%$2d & Upper vibrational quantum number \\
\hline
\end{tabular}
\end{center}
\noindent
*Code definitions:
b2 = $J\p$, $K_a\pp$, $K_c\pp$, $K_a\p$, $K_c\p$, $v_1\pp$, $v_2\pp$, $v_=
3\pp$,
$v_1\p$, $v_2\p$, $v_3\p$;
a5 = $J\p$, $K_a\pp$, $K_c\pp$, $K_a\p$, $K_c\p$;
a1 = $J\p$;
a0 = none further quantum numbers.
\end{table}

\subsection{Dipoles file}\label{dipole_file}

The dipole file mirrors the simple structure of the transitions file,
see Table~\ref{tab:dips}. Two pointers, $i$ and $f$ point to rows in
the \texttt{.states} file to identify the upper and lower states
involved, as specified by the state ID variable, plus the full
information characterising these states.  $D$ gives the signed
transition dipole (in Debye). In this form the combination of the
\texttt{.states} and \texttt{.dipole} files can be used to simulate the
effect of weak linear polarized electric (Stark) fields on the
molecule, both static and time-dependent, where the polarization of
the field is along $Z$.  The $|D|$ value is related to the Einstein
$A$ coefficient from the transitions file via a simple transformation:
\begin{equation}
A_{if}  =  \frac{64\times 10^{-36} \pi^4}{3 h} |D|^2 \frac{|D|^2}{(2J^{\prime}} +1)
\tilde\nu_{if}^3
\end{equation}
where $h$ is Planck's constant, and $J^{\prime\prime}$ is the total
angular momentum of the lower state. However the phase of the (potentially
complex)  $D$, which is important
for calculating the Stark effect,
is lost in this transformation.
Therefore, in the dipole file we provide the
actual value of $D$.

The ExoMol format of the \texttt{.states}/\texttt{.dipole} files
combination can then also be used for modelling higher-order field
effects, including optical activity, polarization phenomena, nonlinear
optical properties, and effects of strong fields including the effects
of strong magnetic fields \cite{04Barron.method}. In this case the
matrix elements of the corresponding properties will be stored as
additional columns after the $D$ column. Table~\ref{tab:dipeg} gives
an example of a dipole file for CaO.

\begin{table}
\caption{Specification of the dipole file.}
\label{tab:dips}
\begin{tabular}{llll} \hline
Field & Fortran Format & C Format & Description\\ \hline
$i$ & \texttt{I12} & \texttt{\%12d} & Upper state ID\\
$f$ & \texttt{I12} & \texttt{\%12d} & Lower state ID\\
$D$  & \texttt{F12.8} & \texttt{\%12.8d} & Dipole moment in Debye\\
\hline
\end{tabular}

\noindent
Fortran format: \texttt{(I12,1x,I12,1x,F12.8)}\\
\end{table}

\SaveVerb{term}|40Ca-16O__VBATHY.dipole|
\begin{table}
\caption{Extract from the dipole file for $^{40}$Ca$^{16}$O,
\protect\UseVerb{term}.}
\label{tab:dipeg}
\begin{center}
\begin{tabular}{ccc}
\hline\hline
$i$     &  $f$          &               $\mu_{if}$\\
\hline
          33    &       1   &    0.52209235E+01     \\
          34    &       1   &    0.14271796E+00     \\
          35    &       1   &    0.50123653E-01     \\
          36    &       1   &    0.16344445E-01     \\
          37    &       1   &   -0.48627678E-03     \\
          38    &       1   &   -0.86965402E-03     \\
          39    &       1   &    0.31660861E-03     \\
          40    &       1   &    0.21892847E-03     \\
          41    &       1   &   -0.15794191E-03     \\
\hline
\hline
\end{tabular}

\noindent
 $i$: Upper  state counting number;
$f$:  Lower  state counting number;
 $\mu_{if}$: transition dipole in Debye.
\end{center}
\end{table}

\subsection{Cross section file}

Absorption cross sections are provided by ExoMol in files with the
extension \texttt{.cross}. The two columns of this file are described
in Table~\ref{tab:cross} and are central bin wavenumber ($\tilde{\nu}$, in
$\mathrm{cm^{-1}}$) and cross section value ($\sigma$, in
$\mathrm{cm^2 molec^{-1}}$). The two fields are separated by a single
space. The stored cross sections may be temperature or temperature and
pressure dependent as denoted by the header which includes the
temperature in K and the pressure in bar.

The absorption cross section is given as a sequence of average values,
$\sigma_i$ within wavenumber bins, $\tilde{\nu}_i$ of specified width,
$\Delta\tilde{\nu}$. That is, in general, cross section values are given by
\begin{equation}
\sigma_i = \sum_j \frac{S_j}{\Delta\tilde{\nu}} \int_{\tilde{\nu}_i -
\Delta\tilde{\nu}/2}^{\tilde{\nu}_i + \Delta\tilde{\nu}/2} f(\tilde{\nu};
\tilde{\nu}_{0,j}\cdots)\,\mathrm{d}\tilde{\nu},
\end{equation}
where the sum is taken over individual absorption lines of strength
$S_j$ and central wavenumber $\tilde{\nu}_{0,j}$, and $f(\tilde{\nu};
\tilde{\nu}_{0,j}\cdots)$ is a normalized lineshape function which is
calculated using transition-dependent parameters,
$(\tilde{\nu}_{0,j}\cdots)$. In the case that the bin width is much
smaller than the line-width, $\Delta\tilde{\nu} \ll \mathrm{HWHM}$,
this formula reduces to
\begin{equation}
\sigma_i = \sum_j S_j f(\tilde{\nu}; \tilde{\nu}_{0,j}\cdots).
\end{equation}
See Hill \ea\ \cite{jt542} for further details.
Table \ref{tab:cross_example} gives an example of a cross section file for
water.

\begin{table}
\caption{Specification of the \texttt{.cross} cross section file
format}\label{tab:cross}
\begin{tabular}{llll}
\hline
Field & Fortran Format & C Format & Description\\
\hline
$\tilde{\nu}_i$ & \texttt{F12.6} & \texttt{\%12.6f} & Central bin wavenumber,
$\mathrm{cm^{-1}}$\\
$\sigma_i$ & \texttt{ES14.8} & \texttt{\%14.8e} & Absorption cross section,
$\mathrm{cm^2molec^{-1}}$\\
\hline
\end{tabular}

\noindent
Fortran format: \texttt{(F12.6,1x,ES14.8)}\\
\end{table}

\SaveVerb{term}|1H2-16O__BT2__0-30000__296K__0bar__0.01.cross|
\begin{table}
\caption{Extract from \protect\UseVerb{term}: Cross section function file for
H$_2{}^{16}$O.}
\label{tab:cross_example} 
\begin{tabular}{ll}
\hline
$\tilde{\nu}$/cm$^{-1}$ & $\sigma$/$\mathrm{cm^2 molec^{-1}}$\\
\hline
 1525.530000 &  8.58106599E-36  \\
 1525.540000 &  4.67629889E-37  \\
 1525.550000 &  4.53342436E-36  \\
 1525.560000 &  6.55402974E-34  \\
 1525.570000 &  2.08219623E-31  \\
 1525.580000 &  1.90580250E-33  \\
 1525.590000 &  4.18283497E-32  \\
 1525.600000 &  1.30145956e-26  \\
 1525.610000 &  5.31710192E-27  \\
 1525.620000 &  7.96854826E-28  \\
 1525.630000 &  1.04806968E-18  \\
 1525.640000 &  7.80260131E-18  \\
 1525.650000 &  3.18257478E-22  \\
 1525.660000 &  1.27578728E-34  \\
 1525.670000 &  6.09403363E-30  \\
 1525.680000 &  1.50281502E-29  \\
 1525.690000 &  2.79935737E-29  \\
\hline
\end{tabular}
\end{table}

\subsection{$k$-coefficients file}

$k$-coefficients are provided by ExoMol in files with the
extension \texttt{.kcoef}. The format of this file is described
in Table~\ref{tab:kcoef}. The $k$-coefficients are provided in $N_{\rm bins}$
wavelength bins spanning a specified wavelength range
$\lambda_{\rm min}$ to $\lambda_{\rm max}$. The spacing between
the bin centers is $\Delta\lambda$ while each bin is $\delta\lambda$
wide. Clearly, for the case that the bins do not overlap
 $\Delta\lambda =\delta\lambda$.
Each bin contains $N_g$ values of $k$ corresponding to the
$g$-ordinate values (which will be an increasing sequence of $k$-coefficients);
the corresponding Gaussian weights, $w$, are also given.

Figure \ref{tab:kcoef_example} gives an example of a $k$-coefficient file for
water.

\begin{table}
\caption{Specification of the \texttt{.kcoef} cross section file format;
the $k$-coefficients are in cm$^2$ molecule$^{-1}$.}\label{tab:kcoef}
\begin{tabular}{llll}
\hline
Field & Fortran Format & C Format & Description\\
\hline
$P$ & \texttt{F12.6} & \texttt{\%12.6f} & Pressure in bar\\
$T$ & \texttt{F8.2} & \texttt{\%8.2f} & Temperature in K\\
$\lambda_{\rm min}$  & \texttt{F12.6} & \texttt{\%12.6f} & Minimum central
wavelength of bins, in $\mu$m,\\
$\lambda_{\rm max}$  & \texttt{F12.6} & \texttt{\%12.6f} & Maximum central
wavelength of bins, in $\mu$m,\\
$\Delta\lambda$      & \texttt{F12.6} & \texttt{\%12.6f} & Spacing of wavelength
bin centres, in $\mu$m,\\
$\delta\lambda$      & \texttt{F12.6} & \texttt{\%12.6f} & Bin widths,
in $\mu$m,\\
$N_{\rm bins}$       & \texttt{I6} & \texttt{\%6d} & No. of wavelength bins
 provided\\
$N_g$                & \texttt{I6} & \texttt{\%6d} & No. of $g$-ordinate values at which $k$ is
provided within each bin\\
$g_{\rm ord}$        & $N_g$\texttt{*(F10.8, 1X)} & $N_g$\texttt{*\%10.8f} &
$N_g$ $g$-ordinate values, $0\leq g\leq 1$ to be read on a single line.\\
$w$        & $N_g$\texttt{*(F10.8, 1X)} & $N_g$\texttt{*\%10.8f} & $N_g$ quadrature weight values to be read on a single line.\\\hline
\end{tabular}
\end{table}

\SaveVerb{term}|1H2-16O__BT2__0-30000__296K__0.01.kcoef|
\begin{longtable}{ll}
\hline                                                      
\verb!     0.005000         !\\
\verb!    240.00      !\\
\verb!      0.300000                                                                      !\\
\verb!     30.000000    !\\
\verb!      0.025000     !\\
\verb!      0.025000                                                                      !\\
\verb!    1189        !\\
\verb!      20    !\\
\verb!  0.00343570 0.01801404 0.04388279 ... 0.98198600 0.99656430     !\\
\verb!  0.0088070035 0.0203007140 0.0313360240 0.0416383710 ...  0.0088070035             !\\
\verb!           0.3 6.41554300e-23 6.42944800e-23 6.60117000e-23 ... 6.32264600e-21      !\\
\verb!        0.3025 6.67216472e-23 6.68662592e-23 6.86521680e-23 ... 7.09400881e-21      !\\
\verb!  ...      !\\
\verb!     30.000000 1.47557489e-24 1.47877304e-24 1.51826910e-24 ... 1.45420858e-22      !\\
\hline                                                                     
               
\caption{Extract from \protect\UseVerb{term}: $k$-coefficient file for                H$_2{}^{16}$O.}\label{tab:kcoef_example}                           
\end{longtable}

\clearpage                                                                
                    
\subsection{Partition function file}               

Temperature-dependent properties such as the partition function and
the cooling function are provided by the partition function file with extension
\texttt{.pf}.  This file gives partition function, $Q$, and, if
available ($C_{\rm avail}=1)$, cooling function, $W$, values as a
function of temperature, $T$. Usually $T$ is given in steps of $T_{\rm
  Step}$ (usually 1 K) up to $T^Q_{\rm max}$. Table~\ref{tab:pf} gives
the specification for the \texttt{.pf} file while Table \ref{tab:pfeg} gives
an example for the CaH molecule.

ExoMol, like HITRAN \cite{HITRAN-A}, follows the convention that the
nuclear spin factors are derived from the full atomic nuclear spin
degeneracies. Thus, for example, for H$_2{}^{17}$O, H has spin
$\frac{1}{2}$ so has degeneracy $(2I_H+1) = 2$ and $^{17}$O has spin $\frac{5}{2}$ so
$(2I_O+1)=6$. This means that para water has degenercy 1 and ortho water has
degeneracy 18. This is in contrast to standard compilations of
astronomical partition functions such as those of Irwin
\cite{81Irwin.partfunc}, Sauval and Tatum
\cite{84SaTaxx.partfunc}, and Barklem and Collet \cite{16BaCoxx.partfunc}
who normalise to unit atomic nuclear spin
degeneracy factors. Final results are independent of which convention
is chosen provided they are used self-consistently.

\begin{table}
\caption{Specification of the \texttt{.pf}  partition function file.}
\label{tab:pf}
\begin{tabular}{llll}
\hline
Field & Fortran Format & C Format & Description\\
\hline
$T$ & \texttt{F8.1} & \texttt{\%8.1f} & Temperature in K\\
$Q(T)$ & \texttt{F15.4} & \texttt{\%15.4f} & Partition function
(dimensionless).\\
$W(T)$ & \texttt{ES12.4} & \texttt{\%12.4e} & Cooling function in ergs s$^{-1}$
molecule$^{-1}$ (if available).\\
\hline
\end{tabular}

\noindent
Fortran format: \texttt{(F8.1,1x,F15.4,1x,ES12.4)} or \texttt{(F8.1,1x,F15.4)}
\\
\end{table}

\SaveVerb{term}|40Ca-1H__Yadin.pf|
\begin{table}
\caption{Extract from the partition function file for $^{40}$Ca$^1$H,
\protect\UseVerb{term}, including cooling function, $W(T)$.}
\label{tab:pfeg} 
\begin{tabular}{rrr}
\hline
$T$/K & $Q(T)$&$W(T)$\\
\hline
      1.0   &       4.0001   &        1.5662E-24   \\
      2.0   &       4.0274   &        6.9307E-22   \\
      3.0   &       4.2080   &        5.1097E-21   \\
      4.0   &       4.5752   &        1.3731E-20   \\
      5.0   &       5.0664   &        2.6214E-20   \\
      6.0   &       5.6252   &        4.3743E-20   \\
      7.0   &       6.2198   &        6.7576E-20   \\
      8.0   &       6.8342   &        9.8328E-20   \\
      9.0   &       7.4602   &        1.3618E-19   \\
     10.0   &       8.0937   &        1.8114E-19   \\
     11.0   &       8.7323   &        1.0279E-18   \\
     12.0   &       9.3745   &        2.6263E-18   \\
     13.0   &      10.0193   &        5.0095E-18   \\
     14.0   &      10.6663   &        8.1986E-18   \\
     15.0   &      11.3148   &        1.2208E-17   \\
     16.0   &      11.9646   &        1.7050E-17   \\
\hline
\end{tabular}
\end{table}

\section{Utility Programs}


As part of the database, ExoMol provides a number of sample utilities
for manipulating data presented in the ExoMol format. This covers the
most common applications of the line lists, such as computing
absorption/emission cross sections \cite{jt542} or coefficients
(`stick' intensities), partition functions, cooling functions and
lifetimes. These utilities are available as Fortran~95 and Python
programs combined into an \textsc{ExoCross} package. ExoCross allows
users to utilise different line profiles to generate cross-sections,
including the Voigt profile, using the line parameters compiled in the
ExoMol database (file \texttt{.broad}) when available. Other standard
line profiles include the temperature-dependent Doppler profile,
rectangular-shape profiles (i.e. averaged cross-sections over the
wavenumber bin), as well as general Gaussian and Lorentzian profiles
with line-widths as free input parameters. The utilities can be
obtained directly from the ExoMol website together with examples of
input files.

As mentioned above a Python utility, \texttt{extract\_\_trans.py}
is provided which reads the \texttt{.trans} in \texttt{.bz2} format
without requiring it to be uncompressed.

The molecular data produced by ExoMol is increasingly being used
in a wide variety of applications. In many of these applications,
pre-existing software packages utilise different data formats for
their input files. The detailed specification of our format above
should be sufficient to allow conversion to any required format using
a suitable parser program.  We have written a number of
utility programs (in Python) that convert the ExoMol format to
other common formats. These can be used to produce the full line
list in different formats such as HITRAN, for which the
Python utility \texttt{exomol2hitran.py} is available.
Oscillator strengths can be generated by the
 utility exomol2gf.py.
However, we should stress that the ExoMol format is much
more compact than most other alternatives (notably HITRAN format),
primarily due to the separation of information into the states and
trans files. The size of the ExoMol line lists means that it can be
prohibitive to store the full data in another format; for example, the
methane 10to10 line list would require 14 TB to be stored in HITRAN
format. It is therefore recommended that the utility programs provided
be integrated within the larger program, such that the new format data
is produced for a small number of lines at a time which are immediately
utilised by the program, thus avoiding the storage of data. 

\section{Website}

The ExoMol website (www.exomol.com) is the main source of the ExoMol
data. Data can be accessed two ways: (a) by  first selecting  a molecule
and then a specific isotopologue-dataset name combination or
(b) by selecting a particular data type, such as partition function.
We note that cross sections are also being made available via
the Virtual Atomic and Molecular Data Centre (VAMDC)
\cite{jt481,jt630}.  This API takes the form of a ``Table Access
Protocol'' service at \texttt{www.exomol.com/tap/sync}.

The exomol.all file described in Section 5.2 is available at
www.exomol.com/exomol.all and the data files comprising each dataset
can be accessed at a URL of the form
\texttt{www.exomol.com/db/<MolFormula>/<IsoSlug>/<DSName>/<Filename>},
for example
\verb!www.exomol.com/db/SiO/28Si-16O/EBJT/28Si-16O__EBJT.pf!.

A separate API for machine-access of the ExoMol data in the format
described in this paper will be available at \texttt{www.exomol.com/api} and
documented on the website. This will allow the automatic retrieval of
data files by the \texttt{HTTP GET} method.

\section{Conclusions}

The ExoMol database presented here is a molecule-by-molecule set of
comprehensive line lists for modelling spectra and other properties of
hot gases. The choice of molecules is dictated by the need to model
the atmospheres of exoplanets and other hot astronomical objects, but
the spectroscopic data have much wider applications than this. We are
still in the process of adding molecules to the database and are
receptive to suggestions of other key species to include.  In
particular the discovery \cite{11LeOdKu.exo,11RoDeDe.exo} and
observation of spectra \cite{jt629} of very hot, rocky planets will
undoubtedly lead to the consideration of novel species
\cite{12ScLoFe.exo,15ItIkMa.exo}.

In this paper we have significantly expanded the flexibility and scope
of the data model used by ExoMol. The original ExoMol format of a
\texttt{.states} and \texttt{.trans} file has been already adopted as
a possible input format by some programs, such as PGOPHER
\cite{PGOPHER}. The extended format proposed here is both
backward-compatible and designed to allow direct extraction of data
from the database using an application programming interface (API).
We believe that this will be helpful to scientists and relieve them of
the need to manually deal with datasets containing many billions of
spectral lines.  Our new data structure allows the inclusion of
important new molecular parameters including those to determine
pressure broadening, state-resolved radiative lifetimes and Land\'e
$g$-factors.

\section*{Acknowledgments}

This work was supported by the ERC under the Advanced Investigator Project
267219. We thank Hengying Li for help preparing the ExoMol website and
Yixin Wang for pointing out the corrections for this update.

\bibliographystyle{model1a-num-names}


\end{document}